\documentclass[11pt]{article}
\usepackage{longtable}
\usepackage{amsmath}
\usepackage{amsfonts}
\usepackage{amssymb}
\usepackage{graphicx}
\usepackage{float}
\newcommand{\barr}[1]{\not\mathrel #1}
\newcommand{\lanl}{\langle}
\newcommand{\ranl}{\rangle}
\newcommand{\be}{\begin{equation}}
\newcommand{\ee}{\end{equation}}
\newcommand{\bea}{\begin{eqnarray}}
\newcommand{\eea}{\end{eqnarray}}

\newcommand{\nn}{\nonumber \\}

\newcommand{\beq}{\begin{equation}}
\newcommand{\eeq}{\end{equation}}
\newcommand{\beqa}{\begin{eqnarray}}
\newcommand{\eeqa}{\end{eqnarray}}
\newcommand{\beqan}{\begin{eqnarray*}}
\newcommand{\eeqan}{\end{eqnarray*}}
\newcommand{\ba}{\begin{array}}
\newcommand{\ea}{\end{array}}

\newcommand{\no}{\nonumber}
\newcommand{\bdm}{\begin{displaymath}}
\newcommand{\edm}{\end{displaymath}}

\begin{document}

{\small \hfill FZJ-IKP(TH)-2000-04}

\bigskip\bigskip
\bigskip\bigskip

\thispagestyle{empty}

\begin{center}

{{\Large\bf 
The chiral effective pion--nucleon Lagrangian of order {\boldmath $p^4$}
\footnote{Work supported in part by Deutsche Forschungsgemeinschaft,
contract Nos. Me~864/11-1 and Schu~439/10-1, and by VEGA grant No.~1/4301/97.}
}}

\end{center}

\vspace{.4in}

\begin{center}
{\large
Nadia Fettes$^a$~\footnote{Present address: Kellogg Radiation
  Laboratory, California Institute of Technology, 
 Pasadena, \hbox{CA 91125}, USA},
Ulf-G. Mei{\ss}ner$^a$, 
Martin Moj\v zi\v s$^b$,
Sven Steininger$^a$~\footnote{Present address: McKinsey Consulting, K\"oln, Germany}
}

\vskip 1cm

{\em $^a$Forschungszentrum J\"ulich, Institut f\"ur Kernphysik (Theorie)}\\
{\em D-52425 J\"ulich, Germany}

\vskip 0.5cm

{\em $^b$Department of Theoretical Physics, Comenius University}\\ 
{\em SK-28415 Bratislava, Slovakia}

\end{center}

\vskip 2cm

\begin{abstract}\noindent
We construct the minimal effective
chiral pion--nucleon SU(2)  Lagrangian at fourth order in the chiral
expansion. The Lagrangian contains 118 in principle measurable terms.
We develop both the relativistic
as well as the heavy baryon formulation of the effective field theory.
For the latter, we also work out explicitly all $1/m$ corrections at
fourth order. We display all relevant relations needed to find the
linearly independent terms.
\end{abstract}

\vfill


\section{Introduction}\label{sec:intro} 
\def\theequation{\arabic{section}.\arabic{equation}}
\setcounter{equation}{0}
Low-energy pion and pion--nucleon physics is described in QCD via chiral perturbation
theory (CHPT)~\cite{GL,GSS}. This is an effective field theory designed to solve
the Ward identities of the chiral symmetry of QCD order by order. CHPT is based
on an effective Lagrangian constructed in a systematic way in terms of
the asymptotically observed hadronic fields and consistent with all
symmetries of QCD.
The effective Lagrangian is organized in the form of a chiral expansion, i.e.
an expansion in powers of momenta and light quark
masses~\cite{wein79}. At every order, the
effective Lagrangian contains free parameters, the so--called low-energy
constants (LECs). Various quantities calculated from the effective Lagrangian
are related by the same set of LECs entering the results. In addition,
there are some rare cases where, to a certain given order in the chiral
expansion, predictions are free of LECs.
At lowest order, CHPT just reproduces the well--known results of 
current algebra. However, within CHPT a systematic improvement of 
current algebra predictions is possible.
Furthermore, to restore unitarity perturbatively, one has to calculate
(pion) loop corrections because the tree diagrams contributing at lowest order are
always real.  Every loop increases the chiral order by two, 
therefore the lowest order results are the tree level ones. 
The relation between current algebra and the lowest order tree level effective
Lagrangian has already been established in the sixties. Loops
enter at higher orders, together with tree contributions with insertions from the
higher order Lagrangians. In the meson sector, the chiral expansion thus proceeds in steps
of two, if one assumes the standard scenario of chiral symmetry breaking
with a large quark--antiquark condensate.

\medskip \noindent
In the one--nucleon sector of CHPT the situation is different. Here,
due to the fermionic nature of the matter fields, couplings with odd
powers in momenta are allowed and the lowest order tree graphs stem
from a dimension one chiral Lagrangian. Loops start to enter at third
order (in a scheme that respects the power counting, see
below). However, for various reasons (convergence, completeness, and
so on) calculations are now performed at the
complete one--loop level, i.e. at fourth chiral order (for a review
and status report,
see refs.\cite{BKMrev,ulfhugs}). The full effective
Lagrangian is, however, only known up to third 
order~\cite{GSS,Krause,BKKM,EM,FMS98}. 
The most general fourth order $\pi$N chiral Lagrangian
is still the missing ingredient in a complete one--loop analysis within
pion--nucleon CHPT. It is the purpose of this paper to provide this 
missing component. We refrain here from discussing the three--flavor
case, which certainly deserves more study in the future.

\medskip \noindent
Most  CHPT calculations for the $\pi$N system (coupled to external
 fields) have been performed in the
framework of Heavy Baryon CHPT (HBCHPT), a specific non-relativistic
projection of the theory. This scheme was developed in ref.\cite{JM}
motivated by the methods used in heavy quark effective field theory
and the observation that a relativistic formulation with baryons leads
to complications in the power counting when standard dimensional
regularization is employed~\cite{GSS}. In HBCHPT the troublesome mass
scale, the nucleon mass $m$, appearing in the fermion propagator, is
simply transformed in a string of vertex corrections with fixed
coefficients and increasing powers in $1/m$. However, as was pointed
out recently in ref.\cite{ET} and ref.\cite{BL}, a scheme consistent
with the power counting, using the relativistic version of the $\pi$N
Lagrangian, can also be set up if one performs a different method of
regularization. Therefore not only the heavy baryon (HB) projection of
the Lagrangian is of interest, but also its fully relativistic version. 
Note also that if one wants to match
the HB approach to the relativistic theory, this is most easily and 
naturally done starting from the relativistic approach (for a detailed
discussion, see e.g.\ ref.\cite{BKKM}). 

\medskip \noindent
The divergent part of the fourth order HBCHPT $\pi$N Lagrangian is
already known \cite{MMS98}. Also, the complete but not minimal fourth
order Lagrangian with virtual photons has been worked out in ref.\cite{MM}.
In addition, recently an attempt to work
out the complete fourth--order isospin--symmetric Lagrangian in the
absence of  external fields has been presented in 
ref.\cite{M99}.\footnote{We will comment in more detail on that
  paper below.} However, the complete fourth order heavy baryon as well as
relativistic $\pi$N Lagrangian (including external fields and in
particular strong isospin breaking) is constructed here for the first time.
The number of LECs in the resulting  Lagrangian turns out to be
larger than 100. This is similar to the case of the two--loop ${\cal O}(p^6)$
meson Lagrangian, see refs.\cite{FS,BCE}.
In case of such a large number of free parameters in the
theory, one might question the predictive power of the whole
approach. However, most physical processes are only sensitive to
a small number of LECs. As an example, we mention
elastic--pion nucleon scattering. At second, third and fourth order,
one has four, four and five independent (combinations of) LECs, respectively, and
these can be easily determined by a fit to the vast amount of
low--energy pion--nucleon scattering data (this argument is spelled
out in more detail in ref.\cite{pin4}). In photo--nucleon processes,
one usually has even fewer LECs, e.g.\ the HBCHPT expression of the electric dipole
amplitude $E_{0+}$ for neutral pion scattering off nucleons involves
only two combinations of LECs up--to--and--including fourth order.

\medskip\noindent
The paper is organized as follows. In Sect.~2, the construction of the
Lagrangian is described step by step (choice of fields, construction of
invariant monomials, elimination of linearly dependent terms,
heavy baryon projection). The result is presented and
discussed in Sect.~3. Appendix~A contains a more detailed discussion of the
elimination of the so-called equation-of-motion terms
in the relativistic Lagrangian. Appendix~B contains some relevant
formulae for the heavy baryon projection.

\section{Construction of the Lagrangian}
\def\theequation{\arabic{section}.\arabic{equation}}
\setcounter{equation}{0}
\subsection{Choice of fields}

Let us briefly collect the basic ingredients for the construction of the effective
$\pi$N Lagrangian. The underlying Lagrangian is that of QCD with massless $u
$ and $d$ quarks, coupled to external hermitian $2\times2$ matrix--valued fields
$v_{\mu},$ $a_{\mu},$ $s$ and $p$ (vector, axial--vector, scalar and
pseudoscalar, respectively)~\cite{GL}
\begin{equation}
\mathcal{L}=\mathcal{L}_{\text{QCD}}^{0}+\bar q\gamma^{\mu}\left(  v_{\mu
}+\gamma_{5}a_{\mu}\right)  q-\bar q(s-i\gamma_{5}p)q\, ,
\qquad q=\binom ud~.
\end{equation}
With suitably transforming
external fields, this Lagrangian is locally $SU(2)_{L}\times SU(2)_{R}\times
U(1)_{V}$ invariant. The chiral $SU(2)_{L}\times SU(2)_{R}$ symmetry of QCD is 
spontaneously broken down to its vectorial subgroup, $SU(2)_V$. To
simplify life further, we disregard the isoscalar axial currents as
well as the winding number density.\footnote{Note that isoscalar axial
currents only play a role in the discussion of the so--called spin
content of the nucleon. That topic can only be addressed properly in  a
three flavor scheme. For a lucid discussion of the $\theta$ term,
we refer to ref.\cite{AS}.} Explicit chiral symmetry breaking, i.e. the
nonvanishing $u$ and $d$ current quark mass, is taken into account by setting
$s=\mathcal{M}=\operatorname*{diag}(m_{u},m_{d}).$

\medskip\noindent
On the level of the effective field theory, the spontaneously broken
chiral symmetry is non-linearly realized in terms of the pion and the nucleon
fields~\cite{CCWZ}. The pion fields $\Phi,$ being coordinates of the chiral
coset space, are naturally represented by elements $u(\Phi)$ of this coset
space. The most convenient choice of fields for the construction of the
effective Lagrangian is given by
\begin{eqnarray}%
u_{\mu} & = & i\{u^{\dagger}(\partial_{\mu}-ir_{\mu})u-u(\partial_{\mu}%
-i\ell_{\mu})u^{\dagger}\} ~,\nn
\chi_{\pm} & = & u^{\dagger}\chi u^{\dagger}\pm u\chi^{\dagger}u~, \nonumber\\
F^{\pm}_{\mu\nu} & = & u^{\dagger}F^{R}_{\mu\nu}u \pm uF^{L}_{\mu\nu}u^{\dagger}  
~, 
\label{BB}%
\end{eqnarray}
where
\begin{eqnarray}
\chi & = & 2B(s+ip)~, \nonumber \\
F_{R}^{\mu\nu} & = & \partial^{\mu}r^{\nu}-\partial^{\nu}r^{\mu}-i[r^{\mu
},r^{\nu}]~, \quad r_{\mu} = v_{\mu}+a_{\mu}~, \nonumber \\
F_{L}^{\mu\nu} & = & \partial^{\mu}\ell^{\nu}-\partial^{\nu}\ell^{\mu}%
-i[\ell^{\mu},\ell^{\nu}]~, \quad \ell_{\mu}  =  v_{\mu}-a_{\mu}~,
\end{eqnarray}
and $B$ is the parameter of the meson Lagrangian of $O(p^{2})$ related
to the strength of the quark--antiquark condensate~\cite{GL}. We work
here in the standard framework, $B \gg F_\pi$, with $F_\pi$ the pion
decay constant. For a discussion of the generalized scenario in the
presence of matter fields, see e.g.\ ref.\cite{BK98}.
The reason why the fields in eq.(\ref{BB}) are so convenient is that they all
transform in the same way under chiral transformations, namely as
\begin{equation}
X\overset{g}{\rightarrow}h(g,\Phi)Xh^{-1}(g,\Phi)~,\label{hXh}%
\end{equation}
where $g$ is an element of  $SU(2)_{L}\times SU(2)_{R}$ and 
the so-called compensator $h(g,\Phi)$ defines a non--linear realization
of the chiral symmetry. The compensator $h(g,\Phi)$ depends on $g$ and $\Phi$
in a complicated way, but since the nucleon field $\Psi$ transforms as
\begin{equation}
\Psi\overset{g}{\rightarrow}h(g,\Phi)\Psi~,\qquad\quad\bar{\Psi}\overset
{g}{\rightarrow}\bar{\Psi}h^{-1}(g,\Phi)\label{hPsi}%
\end{equation}
one can easily construct invariants of the form $\bar{\Psi}O\Psi$ without
explicit knowledge of the compensator. Moreover, for the covariant derivative
defined by
\begin{equation}%
D_{\mu}  =  \partial_{\mu}+\Gamma_{\mu}~, \qquad
\Gamma_{\mu} =  \frac{1}{2}\{u^{\dagger}(\partial_{\mu}-ir_{\mu
})u+u(\partial_{\mu}-i\ell_{\mu})u^{\dagger}\}~,
\label{covD}%
\end{equation}
it follows that also $\left[  D_{\mu},X\right]  ,$ 
$\left[  D_{\mu},\left[  D_{\nu},X\right]
\right]  ,$ etc. transform according to eq.(\ref{hXh}) and $\ D_{\mu}\Psi,$
$D_{\mu}D_{\nu}\Psi,$ etc. according to eq.(\ref{hPsi}). This allows for a simple
construction of invariants containing these derivatives.
The definition of the covariant derivative eq.(\ref{covD}) implies two important
relations. The first one is the so-called curvature relation
\begin{equation}
\left[  D_{\mu},D_{\nu}\right]  =\frac{1}{4}\left[  u_{\mu},u_{\nu}\right]
-\frac{i}{2}F_{\mu\nu}^+~,   \label{curv}%
\end{equation}
which allows one to consider products of covariant derivatives only in the
completely symmetrized form (and ignore all the other possibilities). The
second one is (be aware that some authors like e.g. in
refs.\cite{BCE,BK98}, use a different convention which leads to
a minus sign in the definition of $F_{\mu\nu}^-$ as compared to the
one used here)
\begin{equation}
\left[  D_{\mu},u_{\nu}\right]  -\left[  D_{\nu},u_{\mu}\right]  =F_{\mu\nu}^-~,
   \label{Du anti}%
\end{equation}
which in a similar way allows one to consider covariant derivatives of $u_{\mu}$
only in the explicitly symmetrized form in terms of the tensor $h_{\mu\nu}$,
\begin{equation}
h_{\mu\nu}=\left[  D_{\mu},u_{\nu}\right]  +\left[  D_{\nu},u_{\mu}\right]~.
\end{equation}

\medskip\noindent
For the construction of terms, as well as for further phenomenological
applications, it is advantageous to treat isosinglet and isotriplet components
of the external fields separately. We therefore define
\begin{equation}
\widetilde{X}=X-\frac12\langle X\rangle~,
\end{equation}
where $\left\langle \ldots\right\rangle $ stands for the flavor trace, and we
work with the following set of fields: $u_\mu,$ $\widetilde{\chi}_{\pm},$ $\langle
\chi_{\pm}\rangle,$ $\widetilde{F}_{\pm},$ $\langle F_{+}\rangle$ (the trace
$\langle F_{-}\rangle$ is zero because we have omitted the isoscalar
axial current.)

\medskip\noindent
To construct a hermitian effective Lagrangian, which is not only chiral, but
also parity (P) and charge conjugation (C) invariant, 
one needs to know the transformation properties of the
fields under space inversion, charge conjugation and hermitian conjugation.
For the fields under consideration (and their covariant derivatives), hermitian
conjugation  is equal to $\pm$itself and charge conjugation amounts to $\pm
$transposed, where the signs are given (together with the parity) in 
Table~\ref{signs1}. This table also contains the chiral dimensions of the fields.
This is necessary for a systematic construction of the chiral
effective Lagrangian, order by order. Covariant derivatives acting on pion or
external fields count as quantities of first chiral order.
\begin{table}[H] \centering
\begin{tabular}
[c]{|l|c|c|c|c|c|c|}\hline
& $u_{\mu}$ & $\chi_{+}$ & $\chi_{-}$ & $F_{\mu\nu}^{+}$ & $F_{\mu\nu}^{-}$ &
$D_{\mu}$\\\hline
chiral dimension & $1$ & $2$ & $2$ & $2$ & $2$ & $1$\\
parity & $-$ & $+$ & $-$ & $+$ & $-$ & $+$\\
charge conjugation & $+$ & $+$ & $+$ & $-$ & $+$ & $+$\\
hermitian conjugation & $+$ & $+$ & $-$ & $+$ & $+$ & $+$\\\hline
\end{tabular}
\caption{Chiral dimension and transformation properties 
of the basic fields and the covariant derivative acting on the
pion and the external fields.\label{signs1}}%
\end{table}%
\noindent
In what follows, an analogous information for Clifford algebra
elements, the metric $g_{\mu\nu}$  and the totally antisymmetric 
(Levi-Civita) tensor $\varepsilon_{\lambda\mu\nu\rho}$ (for $d=4)$
together with the covariant derivative acting on the nucleon fields will
be needed (for a more detailed discussion, see ref.\cite{FMS98}). 
In this case, hermitian conjugate equals $\pm\gamma^{0}%
\,$(itself)$\,\gamma^{0}$ and charge conjugation amounts to $\pm$transposed,
where the signs are given (together with parity and chiral dimension) in
Table \ref{signs2}. The covariant derivative acting on nucleon fields counts as a
quantity of zeroth chiral order, since the time component of the
derivative gives the nucleon energy, which cannot be considered a small
quantity. However, the combination $\left(i\barr{D}-m\right)  \Psi$ is of first
chiral order. The minus sign for the charge and hermitian conjugation of
$D_{\mu}\Psi$ as well as the chiral dimension of $\gamma_5$
in Table \ref{signs2} is formal and requires some comment,
which will be given below.%
\begin{table}[H] \centering
\begin{tabular}
[c]{|l|l|l|l|l|l|l|l|}\hline
& $\gamma_{5}$ & $\gamma_{\mu}$ & $\gamma_{\mu}\gamma_{5}$ & $\sigma_{\mu\nu}$%
& $g_{\mu\nu}$ & $\varepsilon_{\lambda\mu\nu\rho}$ & $D_{\mu}\Psi$\\\hline
chiral dimension & \multicolumn{1}{|c|}{$1$} & \multicolumn{1}{|c|}{$0$} &
\multicolumn{1}{|c|}{$0$} & \multicolumn{1}{|c|}{$0$} &
\multicolumn{1}{|c|}{$0$} & \multicolumn{1}{|c|}{$0$} &
\multicolumn{1}{|c|}{$0$}\\
parity & \multicolumn{1}{|c|}{$-$} & \multicolumn{1}{|c|}{$+$} &
\multicolumn{1}{|c|}{$-$} & \multicolumn{1}{|c|}{$+$} &
\multicolumn{1}{|c|}{$+$} & \multicolumn{1}{|c|}{$-$} &
\multicolumn{1}{|c|}{$+$}\\
charge conjugation & \multicolumn{1}{|c|}{$+$} & \multicolumn{1}{|c|}{$-$} &
\multicolumn{1}{|c|}{$+$} & \multicolumn{1}{|c|}{$-$} &
\multicolumn{1}{|c|}{$+$} & \multicolumn{1}{|c|}{$+$} &
\multicolumn{1}{|c|}{$-$}\\
hermitian conjugation & \multicolumn{1}{|c|}{$-$} & \multicolumn{1}{|c|}{$+$} &
\multicolumn{1}{|c|}{$+$} & \multicolumn{1}{|c|}{$+$} &
\multicolumn{1}{|c|}{$+$} & \multicolumn{1}{|c|}{$+$} &
\multicolumn{1}{|c|}{$-$}\\\hline
\end{tabular}
\caption{Transformation properties and chiral dimension
of the elements of the Clifford algebra together with the
metric and Levi-Civita tensors as well as the covariant derivative acting
on the nucleon field.\label{signs2}}%
\end{table}%

\subsection{Invariant monomials}

We are now in the position to combine these building blocks to form
invariant monomials.\footnote{Systematic methods to construct these
invariant monomials can also be found in refs.\cite{Krause,BK98}.}
Any invariant monomial in the effective $\pi$N
Lagrangian  is of the generic form
\begin{equation}
\bar{\Psi}A^{\mu\nu\ldots}\Theta_{\mu\nu\ldots}\Psi+\mathrm{h.c.}~.%
\label{monom}%
\end{equation}
Here, the quantity $A^{\mu\nu\ldots}$ is a product 
of pion and/or external fields and their
covariant derivatives. $\Theta_{\mu\nu\ldots}$, on the other hand, is a product
of a Clifford algebra element $\Gamma_{\mu\nu\ldots}$ and a totally
symmetrized product of $n$ covariant derivatives acting on nucleon fields,
$D_{\alpha\beta\ldots\omega}^{n}=\left\{  D_{\alpha},\left\{  D_{\beta
},\left\{  \ldots,D_{\omega}\right\}  \right\}  \right\}  $,%
\begin{equation}
\Theta_{\mu\nu\ldots\alpha\beta\ldots}=\Gamma_{\mu\nu\ldots}D_{\alpha
\beta\ldots}^{n}~.%
\end{equation}
The Clifford algebra elements are understood to be  expanded in the standard basis
($1$, $\gamma_{5}$, $\gamma_{\mu}$, $\gamma_{\mu}\gamma_{5}$,
$\sigma_{\mu\nu}$) and all
the metric and Levi-Civita tensors are included in $\Gamma_{\mu\nu\ldots}$.
Note that Levi-Civita tensors may have some upper indices, which are, however,
contracted with other indices within $\Theta_{\mu\nu\ldots}$.
The structure given in eq.(\ref{monom}) is not the most general Lorentz invariant
structure, it  already obeys some of the restrictions dictated by 
chiral symmetry. The curvature relation eq.(\ref{curv}) manifests itself in the
fact that we only consider symmetrized products of covariant derivatives acting on $\Psi$. Another
feature of eq.(\ref{monom}) is that, except for the $\varepsilon$-tensors, no
two indices of $\Theta_{\mu\nu\ldots}$ are contracted with each other. The
reason for this lies in the fact that at a given chiral order, $\barr{D}\Psi$ can 
always be replaced by $-im\Psi$, since their difference $\left(\barr{D}+im\right)
\Psi$ is of higher order. One can therefore ignore $D_{\mu}\Psi$ contracted
with $\gamma^{\mu},$ $\gamma^{5}\gamma^{\mu}$ and also with $\sigma
^{\lambda\mu}=i\gamma^{\lambda}\gamma^{\mu}-ig^{\lambda\mu}$. The last
relation explains also why $g^{\lambda\mu}\{D_{\lambda},D_{\mu}\}\Psi$ can be
ignored. Other important restrictions on the structure of $\Theta_{\mu
\nu\ldots}$ are discussed in appendix~\ref{app:EOM}.

\medskip\noindent
To get the complete list of terms contributing to
$A^{\mu\nu\ldots}$, one writes down all 
possible products of the pionic and external fields and covariant derivatives
thereof. Every index of $A^{\mu\nu\ldots}$ increases the chiral order by one,
therefore the overall number of indices of $A^{\mu\nu\ldots}$ (as well as
lower indices of $\Theta_{\mu\nu\ldots}$) is constrained by the chiral order
under consideration. Since the matrix fields do not commute, one has to take
all the possible orderings. To get $A^{\mu\nu\ldots}$ with simple
transformation properties under  charge and hermitian conjugation, all the
products are rewritten in terms of commutators and anticommutators. In the
SU(2) case, this is equivalent to decomposing each product of any two
terms into isoscalar and isovector parts, since the only nontrivial product is
that of two traceless matrices, and the isoscalar (isovector) part of this product
is given by an anticommutator (commutator) of the matrices. After such a
rearrangement of products has been performed, the following relations hold:
\begin{equation}
A^{\dagger}=(-1)^{h_{A}}A~,\qquad A^{c}=(-1)^{c_{A}}A^{T}~,%
\end{equation}
where $(-1)^{h_A}$ and $(-1)^{c_A}$ are determined by the signs from 
Table~\ref{signs1} (as products of factors $\pm1$ for every field and covariant
derivative, and an extra factor $-1$ for every commutator). For the Clifford
algebra elements one has similar relations,
\begin{equation}
\Gamma^{\dagger}=(-1)^{h_{\Gamma}}\gamma^{0}\Gamma\gamma^{0}~,\qquad
\Gamma^{c}=(-1)^{c_{\Gamma}}\Gamma^{T}~,%
\end{equation}
where $h_{\Gamma}$ and $c_{\Gamma}$ are determined by the signs from 
Table~\ref{signs2}.

\medskip\noindent
The monomial eq.(\ref{monom}) can now be written as
\begin{equation}
\bar{\Psi}A^{\mu\nu\ldots\alpha\beta\ldots}\Gamma_{\mu\nu\ldots}%
D_{\alpha\beta\ldots}^{n}\Psi+(-1)^{h_{A}+h_{\Gamma}}\bar{\Psi
}\overleftarrow{D}_{\alpha\beta\ldots}^{n}\Gamma_{\mu\nu\ldots}A^{\mu\nu
\ldots\alpha\beta\ldots}\Psi~.\label{monom2}%
\end{equation}
After elimination of total derivatives ($\overleftarrow{D}^{n}\rightarrow
(-1)^{n}D^{n}$) and subsequent use of the Leibniz rule, one obtains (modulo
higher order terms with derivatives acting on $A^{\mu\nu\ldots}$)
\begin{equation}
\bar{\Psi}A^{\mu\nu\ldots\alpha\beta\ldots}\Gamma_{\mu\nu\ldots}%
D_{\alpha\beta\ldots}^{n}\Psi+(-1)^{h_{A}+h_{\Gamma}+n}\bar{\Psi}%
A^{\mu\nu\ldots\alpha\beta\ldots}\Gamma_{\mu\nu\ldots}D_{\alpha\beta\ldots
}^{n}\Psi~.\label{monom3}%
\end{equation}
We see that to a given chiral order, the second term in eq.(\ref{monom2}) either
doubles or cancels the first one. If the later is true, i.e. if $h_{A}%
+h_{\Gamma}+n$ is odd, this term only contributes at
higher orders and can thus be ignored. Consider now charge conjugation acting on the 
remaining  terms of the type given in eq.(\ref{monom3}) (again
modulo higher order terms, with derivatives acting on $A^{\mu\nu\ldots}$)
\begin{equation}
2\bar{\Psi}A^{\mu\nu\ldots\alpha\beta\ldots}\Gamma_{\mu\nu\ldots
}D_{\alpha\beta\ldots}^{n}\Psi+(-1)^{c_{A}+c_{\Gamma}+n}2\bar{\Psi}%
A^{\mu\nu\ldots\alpha\beta\ldots}\Gamma_{\mu\nu\ldots}D_{\alpha\beta\ldots
}^{n} \Psi~.
\end{equation}
By the same reasoning as before, the terms with odd
$c_{A}+c_{\Gamma}+n$  are to be discarded.
The formal minus sign for the charge and hermitian conjugation of $D_{\mu}%
\Psi$ in Table~\ref{signs2}  takes care of this in a simple way. With
this convention, for any $\Theta_{\mu\nu\ldots}$, the two numbers $h_{\Theta}$
and $c_{\Theta}$ are determined by the entries in  Table~\ref{signs2} 
($h_{\Theta}=$ $h_{\Gamma}+n$, $c_{\Theta}=$ $c_{\Gamma}+n$) 
and invariant monomials eq.(\ref{monom}) of a given chiral order 
are obtained if and only if
\begin{equation}
(-1)^{h_{A}+h_{\Theta}}=1~,\qquad (-1)^{c_{A}+c_{\Theta}}=1~.\label{h c}%
\end{equation}

\subsection{Elimination of terms}
\label{sec:eli}
The list of invariant monomials generated by the complete lists of $A^{\mu
\nu\ldots}$ and $\Theta_{\mu\nu\ldots}$ together with the condition (\ref{h c}) 
is still overcomplete. 
It contains linearly dependent terms,
which can be reduced to a minimal set by use of various identities.
In this paragraph, we will discuss the different types of
such identities which allow to eliminate the linearly dependent terms.

\medskip\noindent
First of all, there are general identities, like the cyclic property
of the trace, implying
\begin{equation}
\left\langle a\left[  b,c\right]  \right\rangle =\left\langle c\left[
a,b\right]  \right\rangle~,
\end{equation}
or Schouten's identity
\begin{equation}
\varepsilon^{\lambda\mu\nu\rho}a^{\tau}+\varepsilon^{\mu\nu\rho\tau}%
a^{\lambda}+\varepsilon^{\nu\rho\tau\lambda}a^{\mu}+\varepsilon^{\rho
\tau\lambda\mu}a^{\nu}+\varepsilon^{\tau\lambda\mu\nu}a^{\rho}=0~.
\end{equation}
Another general identity, frequently used in the construction of 
chiral Lagrangians, is provided by the Cayley-Hamilton theorem. 
For 2$\times$2 matrices $a$ and $b$, this theorem just implies
\begin{equation}
\left\{  a,b\right\}  =a\left\langle b\right\rangle +\left\langle
a\right\rangle b+\left\langle ab\right\rangle -\left\langle a\right\rangle
\left\langle b\right\rangle~.
\end{equation}
This was already accounted for by the separate treatment of the traces
and  the traceless
matrices. However, there are other nontrivial identities among products of
traceless matrices. Let us explicitly mention one such identity, which turns
out to reduce the number of independent terms containing four $u$ fields. 
It reads
\begin{equation}
\{  \widetilde{a},\widetilde{b}\,\}  [  \widetilde{a}%
,\widetilde{c}\,]  - \{  \widetilde{a},\widetilde{c}\, \}  
[\widetilde{a},\widetilde{b}\, ]  = \{  \widetilde{a},\widetilde
{a}\, \}  [  \widetilde{b},\widetilde{c}\, ]  -\widetilde
{a} \{  \widetilde{a}, [  \widetilde{b},\widetilde{c}\, ]\, \}~,
\end{equation}
where $\widetilde{a},\widetilde{b},\widetilde{c}$ are traceless 2$\times$2 matrices.

\medskip\noindent
Another set of identities is provided by the curvature relation eq.(\ref{curv}).
In connection with the Bianchi identity for covariant derivatives,
\begin{equation}
\left[  D_{\lambda},\left[  D_{\mu},D_{\nu}\right]  \right]  +\mathrm{cyclic}%
=0~,
\end{equation}
where ``$\mathrm{cyclic}$'' stands for cyclic permutations, it entails
\begin{equation}
\left[  D_{\lambda},F_{\mu\nu}^+ \right]  +\mathrm{cyclic}=\frac{i}{2}\left[
u_{\lambda},F_{\mu\nu}^- \right]  +\mathrm{cyclic}\label{cyclic Fp}~,%
\end{equation}
where we have used the Leibniz rule and eq.(\ref{Du anti}) on the
right--hand--side. On the other hand,
when combined with eq.(\ref{Du anti}), the curvature relation gives
\begin{equation}
\left[  D_{\lambda},F_{\mu\nu}^- \right]  +\mathrm{cyclic}=\frac{i}{2}\left[
u_{\lambda},F_{\mu\nu}^+ \right]  +\mathrm{cyclic}~,\label{cyclic Fm}%
\end{equation}
where we have used the Jacobi identity $[[u_{\lambda},u_{\mu}],u_{\nu
}]+\mathrm{cyclic}=0$ on the right--hand--side.
These relations can be used for the elimination of
(some) terms which contain $\left[  D_{\lambda},F^\pm_{\mu\nu}\right]  $.

\medskip\noindent
Yet another set of identities is based on the equations of motion (EOM) deduced
from the lowest order $\pi\pi$ 
\begin{equation}
\left[  D_{\mu},u^{\mu}\right]  =\frac{i}{2}\widetilde{\chi}_{-}~,
\label{pi eom}%
\end{equation}%
and $\pi$N Lagrangians (its explicit form is given in paragraph~\ref{sec:dim123})%
\begin{eqnarray}
\left(i\barr{D} -m+\frac{1}{2}{g}_{A}\barr{u}\gamma^{5}\right)  \Psi & = &
0~, \\
\bar{\Psi}\left(  i\overleftarrow{\barr{D}}+m-\frac{1}{2}{g}_A
\barr{u}\gamma^{5}\right)  & = & 0~.
\label{Psi eom}%
\end{eqnarray}
Strictly speaking, in these equations $m$ and $g_A$ refer to the
nucleon mass and the axial--vector coupling constant in the SU(2)
chiral limit ($m_u=m_d=0$, $m_s$ fixed). We will not further specify
this but it should be kept in mind.
One can directly use these EOM or (equivalently) perform specific field
redefinitions --- both techniques yield the same result. The pion EOM is used to
get rid of all the terms containing $h_{\;\mu}^{\mu}$, as well as $[D^{\mu
},h_{\mu\nu}]$, which can be eliminated using eq.(\ref{pi eom}) together with
eqs.(\ref{curv}--\ref{Du anti}). The nucleon EOM, the main effect of which is
a remarkable restriction of the structure of $\Theta_{\mu\nu\ldots}$, is
discussed in full generality in appendix~\ref{app:EOM}. Here, we prefer to
follow the procedure adopted in ref.\cite{FMS98} and give only the
specific relations based on partial integrations and the nucleon EOM 
used in the reduction from the overcomplete to the
minimal set of terms. Some of these relations have already appeared in~\cite{FMS98}, 
but for completeness we prefer to show them all. They
read:
\begin{eqnarray} 
\bar{\Psi} A^\mu i\,D_\mu \Psi + {\rm h.c.} & \doteq & 
2 m\,\bar{\Psi} \gamma_\mu A^\mu \Psi~,
\label{R1}
\\[0.3em]
\bar{\Psi} A^{\mu\nu} D_\nu D_\mu \Psi + {\rm h.c.} & \doteq &
-m\left(\bar{\Psi} \gamma_\mu A^{\mu\nu} i\,D_\nu \Psi + {\rm h.c.} \right)~,
\label{R2}
\\[0.3em]
\bar{\Psi} A^{\mu\nu\lambda} i\,D_\lambda D_\nu D_\mu \Psi + {\rm h.c.} 
& \doteq &
m\left(\bar{\Psi} \gamma_\mu A^{\mu\nu\lambda} D_\lambda D_\nu \Psi + 
{\rm h.c.} \right)~,
\label{R3}
\\[0.3em]
\bar{\Psi} \gamma_5 \gamma_\lambda A^{\mu\lambda} i\,D_\mu \Psi + {\rm h.c.} 
& \doteq &
2i\,m \bar{\Psi} \gamma_5 \sigma_{\mu\lambda} A^{\mu\lambda} \Psi \nonumber\\
&+& \left(\bar{\Psi} \gamma_5 \gamma_\mu A^{\mu\lambda} i\,D_\lambda \Psi 
+ {\rm h.c.} \right)~,
\label{R4}
\end{eqnarray} 
\begin{eqnarray}
\bar{\Psi} \gamma_5 \gamma_\lambda A^{\mu\lambda\alpha} D_\alpha D_\mu \Psi +
{\rm h.c.}
& \doteq &
m\left(\bar{\Psi} \gamma_5 \sigma_{\mu\lambda} A^{\mu\lambda\alpha} 
D_\alpha \Psi + {\rm h.c.} \right)
\nonumber\\ & + & 
\left(\bar{\Psi} \gamma_5 \gamma_\mu A^{\mu\lambda\alpha} D_\alpha 
D_\lambda \Psi + {\rm h.c.} \right)~,
\label{R5}
\\[0.3em]
\bar{\Psi} \gamma_5 \gamma_\lambda A^{\mu\lambda\alpha\beta}
i\,D_\beta D_\alpha D_\mu \Psi~, +
{\rm h.c.}
& \doteq &
m\left(i\,\bar{\Psi} \gamma_5 \sigma_{\mu\lambda} 
A^{\mu\lambda\alpha\beta} D_\beta D_\alpha \Psi + {\rm h.c.} \right)
\nonumber\\ & + & 
\left(\bar{\Psi} \gamma_5 \gamma_\mu A^{\mu\lambda\alpha\beta} 
i\,D_\beta D_\alpha D_\lambda \Psi + {\rm h.c.} \right)~,
\label{R5b}
\\[0.3em]
\bar{\Psi} \sigma_{\alpha\beta} A^{\alpha\beta\mu} i\,D_\mu \Psi + {\rm h.c.} 
& \doteq &
-2m \bar{\Psi} \epsilon_{\alpha\beta\mu\nu} \gamma_5 \gamma^\nu 
A^{\alpha\beta\mu} \Psi
- \left(\bar{\Psi} \sigma_{\beta\mu} A^{\alpha\beta\mu}i\, 
D_\alpha \Psi + {\rm h.c.} \right)
\nonumber\\ & + &
\left(\bar{\Psi} \sigma_{\alpha\mu} A^{\alpha\beta\mu} i\,D_\beta \Psi
  + {\rm h.c.} \right)~,
\label{R6}
\\[0.3em]
\bar{\Psi} \sigma_{\alpha\beta} A^{\alpha\beta\nu\mu} D_\nu D_\mu \Psi + {\rm h.c.} 
& \doteq &
m\left(i\,\bar{\Psi} \epsilon_{\alpha\beta\mu\lambda} \gamma_5 
\gamma^\lambda A^{\alpha\beta\nu\mu} D_\nu \Psi + {\rm h.c.}
\right)\nonumber\\ 
& - &
\left(\bar{\Psi} \sigma_{\beta\mu} A^{\alpha\beta\nu\mu} 
D_\nu D_\alpha \Psi + {\rm h.c.} \right)
\nonumber\\ & + &
\left(\bar{\Psi} \sigma_{\alpha\mu} A^{\alpha\beta\nu\mu} D_\nu
  D_\beta \Psi + {\rm h.c.} \right)~,
\label{R6b}
\\[0.3em] 
\bar{\Psi} \gamma_5 \sigma_{\alpha\beta} A^{\alpha\beta\mu} D_\mu \Psi + {\rm
  h.c.} 
& \doteq &
-\left(\bar{\Psi} \gamma_5 \sigma_{\beta\mu} A^{\alpha\beta\mu}
  D_\alpha \Psi + {\rm h.c.} \right)
\nonumber\\ & + & 
\left(\bar{\Psi} \gamma_5 \sigma_{\alpha\mu} A^{\alpha\beta\mu} 
D_\beta \Psi + {\rm h.c.} \right)~,
\label{R7}
\\[0.3em] 
i\,\bar{\Psi} \gamma_5 \sigma_{\alpha\beta} A^{\alpha\beta\nu\mu} 
D_\nu D_\mu \Psi + {\rm  h.c.} 
& \doteq &
 - 
\left(i\,\bar{\Psi} \gamma_5 \sigma_{\beta\mu} A^{\alpha\beta\nu\mu} 
D_\nu D_\alpha \Psi + {\rm h.c.} \right)
\nonumber\\ & + & 
\left(i\,\bar{\Psi} \gamma_5 \sigma_{\alpha\mu} A^{\alpha\beta\nu\mu} 
D_\nu D_\beta \Psi + {\rm h.c.} \right)~,
\label{R7b}
\\[0.3em]
\bar{\Psi} \gamma_\mu [iD^\mu,A] \Psi & \doteq & 
\frac{g_A}{2} \bar{\Psi} \gamma^\mu \gamma_5 [A,u_\mu] \Psi~,
\label{R8}
\\[0.3em]
\bar{\Psi} \gamma_5 \gamma_\mu [iD^\mu,A] \Psi & \doteq & 
-2m \bar{\Psi} \gamma_5 A \Psi  
-\frac{g_A}{2} \bar{\Psi} \gamma^\mu [A,u_\mu] \Psi~,
\label{R9}
\\[0.3em]
\bar{\Psi} \gamma_5 \gamma_\nu [D^\mu,A^\nu] i\,D_\mu \Psi + {\rm h.c.} 
& \doteq &
0~,
\label{td2}
\\[0.3em]
\bar{\Psi}  \epsilon_{\alpha\beta\mu\lambda} \gamma^\lambda 
[D^\nu,A^{\alpha\beta\mu}] i\, D_\nu \Psi + {\rm h.c.} 
& \doteq &
0~,
\label{td3}
\\[0.3em]
\bar{\Psi} [D^\mu,A^\nu] D_\nu D_\mu \Psi + {\rm h.c.} 
& \doteq &
0~,
\label{td1}
\\[0.3em]
\bar{\Psi} \sigma_{\alpha\beta}  [D^\mu,A^{\alpha\beta\nu}] 
D_\nu  D_\mu \Psi + {\rm  h.c.} 
& \doteq &
0~. 
\label{td4}
\end{eqnarray}
Here, the symbol $\doteq$ means equal up to terms of higher order.

\subsection{Heavy baryon projection}

The heavy baryon (HB) projection of the relativistic theory is well documented in
the literature, see e.g.\ refs.\cite{JM,BKKM,EM}. Here, we only collect some
basic definitions and formulae which are needed in the following. For a more
detailed exposition, we refer to the review~\cite{BKMrev}.
The procedure we adopt follows closely the one in ref.\cite{BKKM}, therefore
we only give some steps for completeness. Basically, one 
considers the mass of the nucleon large compared to
the typical external momenta transferred by pions or external probes and writes
the nucleon four--momentum as $p_\mu = m \, v_\mu + \ell_\mu$, $ p^2 =
m^2$, subject to the condition that $v \cdot \ell \ll m$.
Here, $v_\mu$ is the nucleon four--velocity (in the
rest--frame, we have $v_\mu =( 1 , \vec 0 \, )$). 
Note again that  strictly speaking, the nucleon mass appearing here
should be taken at its value in the chiral limit.
Consequently, one can decompose the wavefunction $\Psi (x)$ into velocity
eigenstates
\begin{equation} 
\Psi (x) = \exp [ -i m v \cdot x ] \, [ N_v(x) + h_v(x) ] 
\end{equation}
with 
\begin{equation} 
\barr v \, N_v = N_v \,\, , \quad \barr  v \, h_v = -h_v \,\, , 
\end{equation}
or in terms of velocity projection operators $P_v^\pm$
\begin{equation} 
P_v^+ N_v = N_v \, , \, P_v^- h_v = h_v \, , \quad P_v^\pm =
\frac{1}{2}(1 \pm \barr v \,) \, , \quad P_v^+ + P_v^- = 1\,  . 
\end{equation}
One now eliminates the small component $h_v(x)$ either by using the
equations of motion or path--integral methods.
The Dirac equation for the velocity--dependent
nucleon field $N_v$ takes the form $i v \cdot \partial N_v = 0$ to lowest
order in $1/m$. Apart from the four--velocity, the covariant 
spin--operator $S_\mu$ is of crucial importance in the HB projection.
Its definition and basic properties are:
\begin{equation} 
S_\mu = \frac{i}{2} \gamma_5 \sigma_{\mu \nu} v^\nu \, , \, 
S \cdot v = 0 \, , \, \lbrace S_\mu , S_\nu \rbrace = \frac{1}{2} \left(
v_\mu v_\nu - g_{\mu \nu} \right) \, , \, [S_\mu , S_\nu] = i
\epsilon_{\mu \nu \alpha \beta} v^\alpha S^\beta \,
 \, , \label{spin}
\end{equation}
in the convention $\epsilon^{0123} = -1$.
After this projection, the Lagrangian does not
contain the nucleon mass term any more and also, all Dirac matrices
can be expressed as combinations of $v_\mu$ and $S_\mu$. Therefore,
one can translate all terms of the relativistic Lagrangian
into their heavy fermion counterpieces.
In addition, there are $1/m$ corrections to the various operators. 
These can be worked out  along the lines spelled out in
appendix~A of ref.\cite{BKKM}. The $\pi$N Lagrangian to fourth order 
for the ``light'' components $N_v$ then becomes:
\begin{eqnarray}
{\cal L}_{\pi N} & = & \bar{N}_v \biggl\{ {\cal A}^{(1)} +  {\cal A}^{(2)} 
+ \gamma_0 {\cal B}^{(1)\dagger}\gamma_0 {\cal C}^{-1(0)} {\cal B}^{(1)}\nonumber\\
& + & {\cal A}^{(3)} 
+ \gamma_0 {\cal B}^{(1)\dagger}\gamma_0 {\cal C}^{-1(1)} {\cal B}^{(1)}
+ \gamma_0 {\cal B}^{(2)\dagger}\gamma_0 {\cal C}^{-1(0)} {\cal B}^{(1)}
+ \gamma_0 {\cal B}^{(1)\dagger}\gamma_0 {\cal C}^{-1(0)} {\cal B}^{(2)}\nonumber\\
& + & {\cal A}^{(4)} 
+ \gamma_0 {\cal B}^{(1)\dagger}\gamma_0 {\cal C}^{-1(2)} {\cal B}^{(1)}
+ \gamma_0 {\cal B}^{(2)\dagger}\gamma_0 {\cal C}^{-1(1)} {\cal B}^{(1)}
+ \gamma_0 {\cal B}^{(1)\dagger}\gamma_0 {\cal C}^{-1(1)} {\cal B}^{(2)}\nonumber \\&&
+ \gamma_0 {\cal B}^{(2)\dagger}\gamma_0 {\cal C}^{-1(0)} {\cal B}^{(2)}
+ \gamma_0 {\cal B}^{(3)\dagger}\gamma_0 {\cal C}^{-1(0)} {\cal B}^{(1)}
+ \gamma_0 {\cal B}^{(1)\dagger}\gamma_0 {\cal C}^{-1(0)} {\cal
  B}^{(3)} \biggr\} N_v~, \nonumber \\ &&
\end{eqnarray}
in the standard notation. Explicit expressions for the operators
${\cal A}^{(i)}$, ${\cal B}^{(i)}$, and  ${\cal C}^{-1(i)}$ are
collected in appendix~\ref{app:LHB}.

\subsection{Checks}
To arrive at the final form of the Lagrangian, which contains more
than 100 independent terms, one
has to perform rather lengthy algebraic manipulations. Due to the
large number of terms, these calculations are prone to errors.
To eliminate the possibility of making simple algebraic
errors, we went through the whole calculation  in two independent ways. We
really calculated everything by hand, but simultaneously we have written codes
in \emph{Mathematica} for the construction of the relativistic Lagrangian as well
as for the extraction of its HB projection. The \emph{Mathematica} programs were not
used just to check the hand-made algebraic manipulations step by step, we
have rather used different approaches in the codes wherever possible. The
final coherence of the results obtained in these two different ways is
a very nontrivial cross--check. 

\section{The Lagrangian}%
\label{sec:Lag}
\def\theequation{\arabic{section}.\arabic{equation}}
\setcounter{equation}{0}
The resulting effective Lagrangian is given by  a string of terms with
increasing chiral dimension,
\begin{equation}
\mathcal{L}_{\pi N} = \mathcal{L}_{\pi N}^{(1)} + 
\mathcal{L}_{\pi N}^{(2)} + \mathcal{L}_{\pi N}^{(3)} + 
\mathcal{L}_{\pi N}^{(4)} + \ldots~,
\end{equation} 
where the ellipsis denotes terms of chiral dimension five (or higher).
We will first briefly summarize the first three terms in this
expansion and then display the novel complete and minimal fourth order
Lagrangian. We always give the Lagrangian in the relativistic as well as
in the heavy baryon formulation.

\subsection{Dimension one, two and three}
\label{sec:dim123}

At lowest order, the effective $\pi$N Lagrangian is given in terms
of two parameters, the nucleon mass  and the axial--vector coupling
constant (in the chiral limit). In its relativistic form, it reads
\begin{equation}
\mathcal{L}_{\pi N}^{(1)}=\bar{\Psi}\biggl(  i\barr{D}-m+\frac{g_{A}}%
{2}\barr{u}\gamma_{5}\biggr)  \Psi~.
\end{equation}%
In the HB formulation, the mass term is absent and all Clifford
algebra elements can be expressed in terms of the nucleon four--velocity
and the spin--vector,
\begin{equation}
\widehat{\mathcal{L}}_{\pi N}^{(1)}=\bar{N}_{v}\left(  iv\cdot
D+g_{A}S\cdot u\right)  N_{v}~.%
\end{equation}%

\medskip\noindent
At second order, seven independent terms with LECs appear, so that
the relativistic Lagrangian reads (the explicit form of the various
operators $O_i^{(2)}$ is given in Table~\ref{dim2})
\begin{equation}
\mathcal{L}_{\pi
  N}^{(2)}=\sum_{i=1}^{7}c_{i}\bar{\Psi}O_{i}^{(2)}\Psi~.
\end{equation}%
The LECs $c_i$ are, of course, finite.
The HB projection is straightforward. We work here with the standard
form (see e.g.\ refs.\cite{BKKM,BKMrev}) and do not transform away the
$(v\cdot D)^2$ term from the kinetic energy (as it was done e.g.\ in
ref.\cite{EM}), 
\begin{equation}
\widehat{\mathcal{L}}_{\pi N}^{(2)}   =\frac{1}{2m}\bar{N}_{v}\left(
\left(  v\cdot D\right)  ^{2}-D^{2}-ig_{A}\left\{  S\cdot D,v\cdot u\right\}
\right)  N_{v} +\sum_{i=1}^{7}\widehat{c}_{i}\bar{N}_{v}
\widehat{O}_{i}^{(2)} N_{v}~.
\end{equation}%
Again, the monomials $\widehat{O}_{i}^{(2)}$ are listed in
Table~\ref{dim2} together with the $1/m$ corrections, which some
of these operators receive (this splitting is done mostly to have
an easier handle on estimating LECs via resonance
saturation~\cite{BKMLEC}).
\renewcommand{\arraystretch}{1.1}
\begin{table}[htb] \centering
\begin{tabular}
[c]{|c|c|c|c|}\hline
$i$ & $O_{i}^{(2)}$ & $\widehat{O}_{i}^{(2)}$ & $2 m (\widehat{c}_{i}-c_{i})%
$\\\hline
\multicolumn{1}{|l|}{$1$} & \multicolumn{1}{|l|}{$\left\langle \chi
_{+}\right\rangle $} & \multicolumn{1}{|l|}{$\left\langle \chi_{+}%
\right\rangle $} & \multicolumn{1}{|l|}{$0$}\\
\multicolumn{1}{|l|}{$2$} & \multicolumn{1}{|l|}{$-\frac{1}{8m^{2}%
}\left\langle u_{\mu}u_{\nu}\right\rangle D^{\mu\nu}+{\rm h.c.}$} &
\multicolumn{1}{|l|}{$\frac{1}{2}\left\langle \left(  v\cdot u\right)
^{2}\right\rangle $} & \multicolumn{1}{|l|}{$-\frac{1}{4} g_{A}^{2}$}\\
\multicolumn{1}{|l|}{$3$} & \multicolumn{1}{|l|}{$\frac{1}{2}\left\langle
u\cdot u\right\rangle $} & \multicolumn{1}{|l|}{$\frac{1}{2}\left\langle
u\cdot u\right\rangle $} & \multicolumn{1}{|l|}{$0$}\\
\multicolumn{1}{|l|}{$4$} & \multicolumn{1}{|l|}{$\frac{i}{4}\left[  u_{\mu
},u_{\nu}\right]  \sigma^{\mu\nu}$} & \multicolumn{1}{|l|}{$\frac{1}{2}\left[
S^{\mu},S^{\nu}\right]  \left[  u_{\mu},u_{\nu}\right]  $} &
\multicolumn{1}{|l|}{$\frac{1}{2}$}\\
\multicolumn{1}{|l|}{$5$} & \multicolumn{1}{|l|}{$\widetilde{\chi}_{+}$} &
\multicolumn{1}{|l|}{$\widetilde{\chi}_{+}$} & \multicolumn{1}{|l|}{$0$}\\
\multicolumn{1}{|l|}{$6$} & \multicolumn{1}{|l|}{$\frac{1}{8m}F_{\mu\nu}%
^{+}\sigma^{\mu\nu}$} & \multicolumn{1}{|l|}{$-\frac{i}{4m}\left[  S^{\mu
},S^{\nu}\right]  F_{\mu\nu}^{+}$} & \multicolumn{1}{|l|}{$2 m $}\\
\multicolumn{1}{|l|}{$7$} & \multicolumn{1}{|l|}{$\frac{1}{8m}\left\langle
F_{\mu\nu}^{+}\right\rangle \sigma^{\mu\nu}$} & \multicolumn{1}{|l|}{$-\frac
{i}{4m}\left[  S^{\mu},S^{\nu}\right]  \left\langle F_{\mu\nu}^{+}%
\right\rangle $} & \multicolumn{1}{|l|}{$0$}\\\hline
\end{tabular}
\caption{Independent dimension two operators for the relativistic and
  the HB Lagrangian. The $1/m$ corrections to these operators in the 
  HB formulation are also displayed.\label{dim2}}%
\end{table}%

\medskip\noindent 
At third order, one has 23 independent terms. We follow the notation of
ref.\cite{FMS98} (see also~\cite{EM}),
\begin{equation}
\mathcal{L}_{\pi N}^{(3)}=\sum_{i=1}^{23}d_{i}\, \bar{\Psi}\, 
O_{i}^{(3)}\, \Psi~.
\end{equation}
The LECs $d_i$ decompose into a renormalized scale--dependent and an
infinite (also scale--dependent) part in the standard manner, $d_i =
d_i^r (\lambda ) + \kappa_i/F^2 L(\lambda )$, with $\lambda$ the
scale of dimensional regularization, $L(\lambda)$ as defined in
ref.\cite{GL}, and the $\beta$--functions
$\kappa_i$, first worked out by Ecker~\cite{GE}.
The HB projection including the $1/m$ corrections reads
\begin{equation}
\widehat{\mathcal{L}}_{\pi N}^{(3)} = 
\sum_{i=1}^{23}\widehat{d}_{i}\bar{N}_{v}\widehat{O}_{i}^{(3)} N_{v}
+ \bar{N}_{v}\, \widehat{O}_{\rm fixed}^{(3)}\, N_{v}
+ \bar{N}_{v}\, \widehat{O}_{\rm div}^{(3)}\, N_{v}~,
\end{equation}
with the corresponding operators and $1/m$ corrections collected in 
Table~\ref{dim3} and ${\cal O}_{\rm fixed}^{(3)}$ given
by~\cite{FMS98}
\begin{eqnarray}
\widehat{O}^{(3)}_{\rm fixed} & = &
\frac{g_A}{8m^2}\,[D^\mu,[D_\mu,S\!\cdot\! u]]
-\frac{g_A^2}{32m^2}\,\epsilon^{\mu\nu\alpha\beta}v_\alpha S_\beta
\langle F^-_{\mu\nu} v\!\cdot\!u\rangle
-i\,\frac{1}{4m^2}\left(v\!\cdot\! D\right)^3
\nonumber \\[0.3em]
& - &
\frac{g_A}{4m^2}\,v\cdot\!\!\stackrel{\leftarrow}{D} S\!\cdot\!u\,v\!\cdot\! D
+\frac{1}{8m^2}\left(i\,D^2\, v\!\cdot\! D + \mbox{h.c.}\right)
\nonumber \\[0.3em]
& - &
\frac{g_A}{4m^2}\left(\{S\!\cdot\! D,v\!\cdot\! u\}\,v\!\cdot\! D + \mbox{h.c.}\right)
+\frac{3g_A^2}{64m^2}\left(i\langle(v\!\cdot\!u)^2\rangle\,v\!\cdot\!D + \mbox{h.
c.}\right)
\nonumber \\[0.3em]
& + &
\frac{1}{32m^2}\left(\epsilon^{\mu\nu\alpha\beta}v_\alpha S_\beta
[u_\mu,u_\nu]\,v\!\cdot\! D + \mbox{h.c.}\right)
-\frac{1}{16m^2}\left(i\,\epsilon^{\mu\nu\alpha\beta}v_\alpha S_\beta
\widetilde{F}^+_{\mu\nu} v\!\cdot\! D + \mbox{h.c.}\right)
\nonumber \\[0.3em]
& -& \frac{1}{32m^2}\left(i\,\epsilon^{\mu\nu\alpha\beta}v_\alpha S_\beta
\langle F^+_{\mu\nu}\rangle v\!\cdot\! D + \mbox{h.c.}\right)
\nonumber \\[0.3em]
& - & 
\frac{g_A}{8m^2}\left(S\!\cdot\! u\,D^2 + \mbox{h.c.}\right)
-\frac{g_A}{4m^2}\left(S\cdot\!\!\stackrel{\leftarrow}{D} u\!\cdot\! D + \mbox{h.
c.}\right)
\nonumber \\[0.3em]
& - &
\frac{1+2c_6}{8m^2}\,\left(i\,\epsilon^{\mu\nu\alpha\beta}v_\alpha S_\beta 
\widetilde{F}^+_{\mu\sigma}v^\sigma D_\nu + \mbox{h.c.}\right)
-\frac{1+2 c_6 +4 c_7}{16m^2}\left(i\,\epsilon^{\mu\nu\alpha\beta}v_\alpha S_\beta 
\langle F^+_{\mu\sigma}\rangle\,v^\sigma D_\nu + \mbox{h.c.}\right)
\nonumber \\[0.3em]
& + &
\frac{1+g_A^2+8m\,c_4}{16m^2}\left(\epsilon^{\mu\nu\alpha\beta}v_\alpha S_\beta
[u_\mu,v\!\cdot\! u] D_\nu + \mbox{h.c.}\right)
\nonumber \\[0.3em]
& + &
\frac{g_A}{32m^2}\left(i\,\epsilon^{\mu\nu\alpha\beta} v_\alpha
F^-_{\mu\nu} D_\beta + \mbox{h.c.}\right)
-\frac{g_A^2}{16m^2}\left(i\,v\!\cdot\!u\,u\!\cdot\!D  +
  \mbox{h.c.}\right)
\nonumber \\[0.3em]
& + &
i\,\frac{1+8m\,c_4}{32m^2}\,[v\!\cdot\! u,[D^\mu,u_\mu]]
+\frac{c_2}{2m}\left(i\langle v\!\cdot\!u\,u_\mu\rangle D^\mu + \mbox{h.c.}\right
)~.
\end{eqnarray}
In addition, there are eight terms just needed for the
renormalization (we give these here for completeness). Their explicit form is: 
\begin{eqnarray}
\widehat{O}^{(3)}_{\rm div} 
& = &
\tilde{d}_{24}(\lambda)\,i\left(v\!\cdot\!D\right)^3
+ \tilde{d}_{25}(\lambda)\,v\cdot\!\!\stackrel{\leftarrow}{D}
S\!\cdot\!u\,v\!\cdot\! D
+ \tilde{d}_{26}(\lambda)\left(i\,\langle u\!\cdot\!u\rangle\,v\!\cdot\!D +
  \mbox{h.c.}\right)
\nonumber \\[0.3em]
& + &
\tilde{d}_{27}(\lambda)\left(i\,\langle (v\!\cdot\!u)^2\rangle\,v\!\cdot\!D +
  \mbox{h.c.}\right)
+ \tilde{d}_{28}(\lambda)\left(i\,\langle\chi_+\rangle\,v\!\cdot\!D +
  \mbox{h.c.}\right)
\nonumber \\[0.3em]
& + &
\tilde{d}_{29}(\lambda)\left(S^\mu[v\!\cdot\!D,u_\mu]\,v\!\cdot\!D +
  \mbox{h.c.}\right)
+ \tilde{d}_{30}(\lambda)\left(\epsilon^{\mu\nu\alpha\beta}v_\alpha S_\beta
[u_\mu,u_\nu] \,v\!\cdot\!D + \mbox{h.c.}\right)
\nonumber \\[0.3em]
& + &
\tilde{d}_{31}(\lambda)\left(\epsilon^{\mu\nu\alpha\beta}v_\alpha S_\beta
\tilde{F}^+_{\mu\nu}\,v\!\cdot\!D + \mbox{h.c.}\right)~.
\end{eqnarray}
The LECs $\tilde{d}_i$ have no finite piece~\cite{FMS98}.

\renewcommand{\arraystretch}{1.1}
\begin{table}[H] \centering
\begin{tabular}
[c]{|c|c|c|c|}\hline
$i$ & $O_{i}^{(3)}$ & $\widehat{O}_{i}^{(3)}$ & $4 m^2 (\widehat{d}_{i}-d_{i})%
$\\\hline
\multicolumn{1}{|r|}{$1$} & \multicolumn{1}{|l|}{$-\frac{1}{2m}[u_{\mu
},[D_{\nu},u^{\mu}]]D^{\nu}+{\rm h.c.}$} & \multicolumn{1}{|l|}{$i[u_{\mu},[v\cdot
D,u^{\mu}]]$} & \multicolumn{1}{|l|}{$0$}\\
\multicolumn{1}{|r|}{$2$} & \multicolumn{1}{|l|}{$-\frac{1}{2m}[u_{\mu
},[D^{\mu},u_{\nu}]]D^{\nu}+{\rm h.c.}$} & \multicolumn{1}{|l|}{$i[u_{\mu},[D^{\mu
},v\cdot u]]$} & \multicolumn{1}{|l|}{$-\frac{1}{8}(1+8mc_{4})$}\\
\multicolumn{1}{|r|}{$3$} & \multicolumn{1}{|l|}{$\frac{1}{12m^{3}}[u_{\mu
},[D_{\nu},u_{\rho}]]D^{\mu\nu\rho}+{\rm h.c.}$} & \multicolumn{1}{|l|}{$i[v\cdot
u,[v\cdot D,v\cdot u]]$} & \multicolumn{1}{|l|}{$\frac{1}{8}g_{A}^{2}$}\\
\multicolumn{1}{|r|}{$4$} & \multicolumn{1}{|l|}{$-\frac{1}{2m}\epsilon
^{\mu\nu\alpha\beta}\langle u_{\mu}u_{\nu}u_{\alpha}\rangle D_{\beta}+{\rm h.c.}$} &
\multicolumn{1}{|l|}{$i\epsilon^{\mu\nu\alpha\beta}v_{\beta}\langle u_{\mu
}u_{\nu}u_{\alpha}\rangle$} & \multicolumn{1}{|l|}{$\frac{1}{16}g_{A}$}\\
\multicolumn{1}{|r|}{$5$} & \multicolumn{1}{|l|}{$\frac{1}{2m}i[\chi
_{-},u_{\mu}]D^{\mu}+{\rm h.c.}$} & \multicolumn{1}{|l|}{$[\chi_{-},v\cdot u]$} &
\multicolumn{1}{|l|}{$0$}\\
\multicolumn{1}{|r|}{$6$} & \multicolumn{1}{|l|}{$\frac{1}{2m}i[D^{\mu
},\widetilde{F}_{\mu\nu}^{+}]D^{\nu}+{\rm h.c.}$} & \multicolumn{1}{|l|}{$v^{\nu}[D^{\mu
},\widetilde{F}_{\mu\nu}^{+}]$} & \multicolumn{1}{|l|}{$-\frac{1}{4}(1+2c_{6})$}\\
\multicolumn{1}{|r|}{$7$} & \multicolumn{1}{|l|}{$\frac{1}{2m}i[D^{\mu
},\langle F_{\mu\nu}^{+}\rangle]D^{\nu}+{\rm h.c.}$} & \multicolumn{1}{|l|}{$v^{\nu
}[D^{\mu},\langle F_{\mu\nu}^{+}\rangle]$} & \multicolumn{1}{|l|}{$-\frac{1}{8}
(1+ 2 c_6 + 4 c_{7})$}\\
\multicolumn{1}{|r|}{$8$} & \multicolumn{1}{|l|}{$\frac{1}{2m}i\epsilon
^{\mu\nu\alpha\beta}\langle\widetilde{F}_{\mu\nu}^{+}u_{\alpha}\rangle
D_{\beta}+{\rm h.c.}$} & \multicolumn{1}{|l|}{$\epsilon^{\mu\nu\alpha\beta}v_{\beta
}\langle\widetilde{F}_{\mu\nu}^{+}u_{\alpha}\rangle$} &
\multicolumn{1}{|l|}{$\frac{1}{16}g_{A}$}\\
\multicolumn{1}{|r|}{$9$} & \multicolumn{1}{|l|}{$\frac{1}{2m}i\epsilon
^{\mu\nu\alpha\beta}\langle F_{\mu\nu}^{+}\rangle u_{\alpha}D_{\beta}+{\rm h.c.}$} &
\multicolumn{1}{|l|}{$\epsilon^{\mu\nu\alpha\beta}v_{\beta}\langle F_{\mu\nu
}^{+}\rangle u_{\alpha}$} & \multicolumn{1}{|l|}{$\frac{1}{16}g_{A}$}\\
\multicolumn{1}{|r|}{$10$} & \multicolumn{1}{|l|}{$\frac{1}{2}\gamma^{\mu
}\gamma_{5}\langle u^{2}\rangle u_{\mu}$} & \multicolumn{1}{|l|}{$S\cdot u
\langle u^{2}\rangle $} & \multicolumn{1}{|l|}{$0$}\\
\multicolumn{1}{|r|}{$11$} & \multicolumn{1}{|l|}{$\frac{1}{2}\gamma^{\mu
}\gamma_{5}\langle u_{\mu}u_{\nu}\rangle u^{\nu}$} &
\multicolumn{1}{|l|}{$S^{\mu}u^{\nu}\langle u_{\mu}u_{\nu}\rangle $} &
\multicolumn{1}{|l|}{$0$}\\
\multicolumn{1}{|r|}{$12$} & \multicolumn{1}{|l|}{$-\frac{1}{8m^{2}}%
\gamma^{\mu}\gamma_{5}\langle u_{\lambda}u_{\nu}\rangle u_{\mu}D^{\lambda\nu}+{\rm h.c.}%
$} & \multicolumn{1}{|l|}{$S\cdot u\langle(v\cdot u)^{2}\rangle $} &
\multicolumn{1}{|l|}{$-\frac{1}{2}g_{A}\left(  1+4mc_{4}\right)-\frac{1}{8} g_{A}^{3}$}\\
\multicolumn{1}{|r|}{$13$} & \multicolumn{1}{|l|}{$-\frac{1}{8m^{2}}%
\gamma^{\mu}\gamma_{5}\langle u_{\mu}u_{\nu}\rangle u_{\lambda}D^{\lambda\nu}+{\rm h.c.}%
$} & \multicolumn{1}{|l|}{$\langle S\cdot u\,\,\,v\cdot u\rangle v\cdot u$}%
& \multicolumn{1}{|l|}{$\frac{1}{2}g_{A}\left(  1+4mc_{4}\right)  +\frac{1}{4} g_{A}^{3}$}\\
\multicolumn{1}{|r|}{$14$} & \multicolumn{1}{|l|}{$\frac{1}{4m}i\sigma^{\mu
\nu}\langle\lbrack D_{\lambda},u_{\mu}]u_{\nu}\rangle D^{\lambda}+{\rm h.c.}$} &
\multicolumn{1}{|l|}{$-i[S^{\mu},S^{\nu}]\langle\lbrack v\cdot D,u_{\mu}%
]u_{\nu}\rangle$} & \multicolumn{1}{|l|}{$0$}\\
\multicolumn{1}{|r|}{$15$} & \multicolumn{1}{|l|}{$\frac{1}{4m}i\sigma^{\mu
\nu}\langle u_{\mu}[D_{\nu},u_{\lambda}]\rangle D^{\lambda}+{\rm h.c.}$} &
\multicolumn{1}{|l|}{$-i[S^{\mu},S^{\nu}]\langle u_{\mu}[D_{\nu},v\cdot
u]\rangle$} & \multicolumn{1}{|l|}{$\frac{1}{4} g_{A}^{2}$}\\
\multicolumn{1}{|r|}{$16$} & \multicolumn{1}{|l|}{$\frac{1}{2}\gamma^{\mu
}\gamma_{5}\langle\chi_{+}\rangle u_{\mu}$} & \multicolumn{1}{|l|}{$S\cdot u
\langle\chi_{+}\rangle $} & \multicolumn{1}{|l|}{$0$}\\
\multicolumn{1}{|r|}{$17$} & \multicolumn{1}{|l|}{$\frac{1}{2}\gamma^{\mu
}\gamma_{5}\langle\chi_{+}u_{\mu}\rangle$} & \multicolumn{1}{|l|}{$
\langle S\cdot u\chi_{+}\rangle$} & \multicolumn{1}{|l|}{$0$}\\
\multicolumn{1}{|r|}{$18$} & \multicolumn{1}{|l|}{$\frac{1}{2}\,i\gamma^{\mu
}\gamma_{5}[D_{\mu},\chi_{-}]$} & \multicolumn{1}{|l|}{$\,i [S\cdot D
,\chi_{-}]$} & \multicolumn{1}{|l|}{$0$}\\
\multicolumn{1}{|r|}{$19$} & \multicolumn{1}{|l|}{$\frac{1}{2}i\gamma^{\mu
}\gamma_{5}[D_{\mu},\langle\chi_{-}\rangle]$} & \multicolumn{1}{|l|}{$i
[S\cdot D,\langle\chi_{-}\rangle]$} & \multicolumn{1}{|l|}{$0$}\\
\multicolumn{1}{|r|}{$20$} & \multicolumn{1}{|l|}{$-\frac{1}{8m^{2}}%
i\gamma^{\mu}\gamma_{5}[\widetilde{F}_{\mu\nu}^{+},u_{\lambda}]D^{\lambda\nu}+{\rm h.c.}%
$} & \multicolumn{1}{|l|}{$\,iS^{\mu}v^{\nu}[\widetilde{F}_{\mu\nu}^{+},v\cdot
u]$} & \multicolumn{1}{|l|}{$\frac{1}{2} g_{A}\left(  1+c_{6}\right)$}\\
\multicolumn{1}{|r|}{$21$} & \multicolumn{1}{|l|}{$\frac{1}{2}i\gamma^{\mu
}\gamma_{5}v^{\nu}[\widetilde{F}_{\mu\nu}^{+},u^{\nu}]$} &
\multicolumn{1}{|l|}{$iS^{\mu}[\widetilde{F}_{\mu\nu}^{+},u^{\nu}]$} &
\multicolumn{1}{|l|}{$0$}\\
\multicolumn{1}{|r|}{$22$} & \multicolumn{1}{|l|}{$\frac{1}{2}\,\gamma^{\mu
}\gamma_{5}[D^{\nu},F_{\mu\nu}^{-}]$} & \multicolumn{1}{|l|}{$\,S^{\mu}%
[D^{\nu},F_{\mu\nu}^{-}]$} & \multicolumn{1}{|l|}{$0$}\\
\multicolumn{1}{|r|}{$23$} & \multicolumn{1}{|l|}{$\frac{1}{2}\gamma_{\mu
}\gamma_{5}\epsilon^{\mu\nu\alpha\beta}\langle u_{\nu}F_{\alpha\beta
}^{-}\rangle$} & \multicolumn{1}{|l|}{$\epsilon^{\mu\nu\alpha\beta}
S_{\mu}\langle u_{\nu}F_{\alpha\beta}^{-}\rangle$} &
\multicolumn{1}{|l|}{$0$}\\\hline
\end{tabular}
\caption{Independent dimension three operators for the relativistic and
  the HB Lagrangian. The $1/m$ corrections to these operators in the 
  HB formulation are also displayed.
  Note that we have corrected for two typographical
  errors which appeared in {\protect\cite{FMS98}} in the $1/m$ corrections to
  the operators $\widehat{O}_{4}^{(3)}$ and $\widehat{O}_{20}^{(3)}$. 
\label{dim3}}%
\end{table}%

\subsection{Dimension four}

This paragraph constitutes the main result of our work. We have found
that there are 118 independent dimension four operators, which
are in principle measurable.\footnote{We note that the mastertable
in ref.\cite{MMS98} contains more terms, 199 to be precise. In that
work, however, no attempt was made to find the minimal basis. This is
similar to what was done for the dimension three Lagrangian, compare
e.g.\ refs.\cite{GE} and \cite{FMS98}.} 
The last four ($i=115,\ldots ,118)$, however, are special in the sense
that they are pure contact interactions of the nucleons with the external
sources that have no pion matrix elements. For example, the operators
115 and 116 contribute to the scalar nucleon form factor but not to
pion--nucleon scattering. 
The relativistic dimension four Lagrangian takes the form
\begin{equation}
\mathcal{L}_{\pi N}^{(4)}=\sum_{i=1}^{118} e_{i} \, \bar{\Psi} 
\,O_{i}^{(4)}\, \Psi~,
\end{equation}
and the monomials $O_{i}^{(4)}$ are tabulated in
Table~\ref{dim4}. Also given in that table are 
the processes and observables with the least number of particles 
(including nucleons, pions, photons and W bosons) to which each operator can contribute.
As pointed out in the introduction, many of the operators 
contribute only to very exotic processes, like three or four pion
production induced by photons or pions. Moreover, for a given
process, it often happens that some of these operators appear in
certain linear combinations. Note also that operators with a separate
$\langle \chi_+ \rangle$ simply amount to a quark mass renormalization
of the corresponding dimension two operator.
The HB projection leads to a much more complicated Lagrangian,
\begin{eqnarray}\label{L4HB}
\widehat{{\cal L}}_{\pi N}^{(4)} &=&\bar{N}_{v}\left( \sum_{i=1}^{118}
\hat{e}_{i}\widehat{O}_{i}^{(4)}+\sum_{i=1}^{23}e_{i}^{\prime
}W_{i}+\sum_{i=1}^{67}\left( e_{i}^{\prime \prime }X_{i}^{\lambda
}D_{\lambda }+{\rm h.c.}\right) \right. \nonumber  \\
&&\left. +\sum_{i=1}^{23}\left( e_{i}^{\prime \prime \prime }D_{\mu
}Y_{i}^{\mu \nu }D_\nu +{\rm h.c.}\right) +\sum_{i=1}^{4}\left(
e_{i}^{iv}Z_{i}^{\lambda \mu \nu }D_{\lambda }D_{\mu }D_{\nu }+{\rm h.c.}%
\right) \right.    \nonumber\\
&&\left. +\frac{1}{8m^{3}}\left( v\cdot DD_{\mu }D^{\mu }v\cdot D-\left(
v\cdot D\right) ^{4} \right) +  \widehat{O}_{\rm div}^{(4)}\right) N_{v}~,
\end{eqnarray}
where the first sum contains the 118 low--energy constants
in the basis of the heavy nucleon fields. Various additional terms appear:
First, there are the leading $1/m$ corrections to (most of) the 118
dimension four operators. These contribute to the difference in the
LECs $e_i$ and $\hat{e}_i$; the $\hat{e}_i$ are tabulated in Table~\ref{eim}. Furthermore
we have additional terms with fixed coefficients. They can
be most compactly represented by counting the number of covariant
derivatives acting on the nucleon fields.
The corresponding (tensor) structures $W_i$, $X_i^\lambda$,
$Y_i^{\mu\nu}$, and $Z_i^{\lambda\mu\nu}$ are collected in
Tables~\ref{tab:Wi}, \ref{tab:Xi}, \ref{tab:Yi}, \ref{tab:Zi} 
together with the pertinent 
coefficients $e_i',\ldots,e_i^{iv}$.
There are two other fixed coefficient terms which are listed in the last
line of eq.(\ref{L4HB}).


\renewcommand{\arraystretch}{1.1}
\setlength{\LTcapwidth}{\textwidth}

\begin{longtable}[c]{|c|c|c|c|}  

\hline
$i$ & $O_{i}^{(4)}$ & $\widehat{O}_{i}^{(4)}$ &\\
\hline
\hline
\endhead
\hline
\caption[]{\rule{0cm}{2em}}
\endfoot
\hline
\caption[Independent dimension four operators for the relativistic and
  the HB Lagrangian. Also shown is the physical process with the least
  number of particles to which these operators contribute. Here,
  $N,\pi,\gamma(^*), W$ denotes nucleons, pions, real (virtual)
  external photons and the $W$--boson external sources, respectively.
  Electromagnetic (weak) form factors are abbreviated as ``em. (weak)
  ff.'' and $m$ is the nucleon mass. 
  The last four terms have no pion matrix elememts.]{\label{dim4} 
\rule{0cm}{2em}Independent dimension four operators for the relativistic and
  the HB Lagrangian. Also shown is the physical process with the least
  number of particles to which these operators contribute. Here,
  $N,\pi,\gamma(^*), W$ denote nucleons, pions, real (virtual)
  external photons and the $W$--boson external sources, respectively.
  Electromagnetic (weak) form factors are abbreviated as ``em. (weak)
  ff.'' and $m$ is the nucleon mass. 
  The last four terms have no pion matrix elememts.} 
\endlastfoot
\multicolumn{1}{|r|}{$1$} & \multicolumn{1}{|l|}{$\left\langle u\cdot
u\right\rangle \left\langle u\cdot u\right\rangle $} & \multicolumn{1}{|l|}{$%
\left\langle u\cdot u\right\rangle \left\langle u\cdot u\right\rangle $}
& \multicolumn{1}{|l|}{$\pi N \rightarrow \pi\pi\pi N$}  \\ 
\multicolumn{1}{|r|}{$2$} & \multicolumn{1}{|l|}{$\left\langle u_{\mu
}u_{\nu }\right\rangle \left\langle u^{\mu }u^{\nu }\right\rangle $} & 
\multicolumn{1}{|l|}{$\left\langle u_{\mu }u_{\nu }\right\rangle
\left\langle u^{\mu }u^{\nu }\right\rangle $} 
& \multicolumn{1}{|l|}{$\pi N \rightarrow \pi\pi\pi N$}\\ 
\multicolumn{1}{|r|}{$3$} & \multicolumn{1}{|l|}{$-\frac{1}{4m^{2}}%
\left\langle u\cdot u\right\rangle \left\langle u_{\mu }u_{\nu
}\right\rangle D^{\mu \nu } + {\rm h.c.}$} & \multicolumn{1}{|l|}{$\left\langle u\cdot
u\right\rangle \left\langle (v\cdot u)^{2}\right\rangle $} 
& \multicolumn{1}{|l|}{$\pi N \rightarrow \pi\pi\pi N$}\\ 
\multicolumn{1}{|r|}{$4$} & \multicolumn{1}{|l|}{$-\frac{1}{4m^{2}}%
\left\langle u_{\lambda }u_{\mu }\right\rangle \left\langle u^{\lambda
}u_{\nu }\right\rangle D^{\mu \nu } + {\rm h.c.}$} & \multicolumn{1}{|l|}{$\left\langle
u_{\lambda }v\cdot u\right\rangle \left\langle u^{\lambda }v\cdot
u\right\rangle $} & \multicolumn{1}{|l|}{$\pi N \rightarrow \pi\pi\pi N$}\\ 
\multicolumn{1}{|r|}{$5$} & \multicolumn{1}{|l|}{$\frac{1}{48m^{4}}%
\left\langle u_{\lambda }u_{\mu }\right\rangle \left\langle u_{\nu }u_{\rho
}\right\rangle D^{\lambda \mu \nu \rho } + {\rm h.c.}$} & \multicolumn{1}{|l|}{$%
\left\langle (v\cdot u)^{2}\right\rangle \left\langle (v\cdot
u)^{2}\right\rangle $} & \multicolumn{1}{|l|}{$\pi N \rightarrow \pi\pi\pi N$}\\ 
\multicolumn{1}{|r|}{$6$} & \multicolumn{1}{|l|}{$\frac{i}{2}\left[ u_{\mu
},u_{\nu }\right] \left\langle u\cdot u\right\rangle \sigma ^{\mu \nu }$} & 
\multicolumn{1}{|l|}{$[S^{\mu },S^{\nu }]\left[ u_{\mu },u_{\nu }\right]
\left\langle u\cdot u\right\rangle $}& \multicolumn{1}{|l|}{$\pi N \rightarrow \pi\pi\pi N$} \\
\hline
\end{longtable}

\setlength{\LTcapwidth}{\textwidth}

\begin{longtable}[c]{|c|c|}  
\hline
$i$ & $8m^{3}\left( \hat{e}_{i}-e_{i}\right) $ \\ \hline
\hline
\hline
\endhead
\hline
\caption[]{\rule{0cm}{2em}}
\endfoot
\hline
\caption[$1/m$ corrections to the dimension four LECs $\hat{e}_i$
  in the heavy nucleon basis. Terms which are not subject to such 
  corrections are not listed.]{\label{eim}
\rule{0cm}{2em}$1/m$ corrections to the dimension four LECs $\hat{e}_i$
  in the heavy nucleon basis. Terms which are not subject to such 
  corrections are not listed.}
\endlastfoot 
\multicolumn{1}{|r|}{$1$} & \multicolumn{1}{|l|}{$-\frac{1}{32}g_{A}^{2}+m c_2-%
\frac{1}{8}mc_{4}$} \\ 
\multicolumn{1}{|r|}{$2$} & \multicolumn{1}{|l|}{$\frac{1}{32}g_{A}^{2}-m c_2+%
\frac{1}{8}mc_{4}$} \\ 
\multicolumn{1}{|r|}{$3$} & \multicolumn{1}{|l|}{$\frac{1}{64}\left(
4g_{A}^{2}+g_{A}^{4}\right) -\frac{1}{8}g_{A}^{2}mc_{3}+\frac{1}{4}%
(2+g_{A}^{2})mc_{4}+\left( mc_{4}\right) ^{2}$} \\ 
\multicolumn{1}{|r|}{} & \multicolumn{1}{|l|}{$-g_{A}m^{2}d_{10}+ 2 m^2 \left(d_{14}-d_{15}\right) $} \\ 
\multicolumn{1}{|r|}{$4$} & \multicolumn{1}{|l|}{$-\frac{1}{16}g_{A}^{2}-%
\frac{1}{4}\left( 2+ g_{A}^{2}\right) mc_{4}-\left( mc_{4}\right)
^{2}-g_{A}m^{2} d_{11}-2 m^2 \left( d_{14}-d_{15}\right) $} \\ 
\multicolumn{1}{|r|}{$5$} & \multicolumn{1}{|l|}{$-\frac{1}{64}g_{A}^{4}-%
\frac{1}{8}g_{A}^{2}mc_{2}-g_{A}m^{2}\left( d_{12}+d_{13}\right) $} \\ 
\multicolumn{1}{|r|}{$6$} & \multicolumn{1}{|l|}{$-\frac{1}{32}g_{A}^{2}+%
\frac{1}{4}mc_{3}$} \\ 
\multicolumn{1}{|r|}{$7$} & \multicolumn{1}{|l|}{$\frac{1}{32}\left(
g_{A}^{2}+g_{A}^{4}\right) +\frac{1}{4}mc_{2}+\frac{3}{8}g_{A}^{2}mc_{4}$}\\
\hline
\end{longtable}

\renewcommand{\arraystretch}{1.2}
\setlength{\LTcapwidth}{\textwidth}

\begin{longtable}[c]{|c|c|c|}  
\hline
$i$ & $W_{i}$ & $8m^{3}e_{i}^{\prime }$ \\ \hline
\hline
\hline
\endhead
\hline
\caption[]{\rule{0cm}{2em}}
\endfoot
\hline
\caption[Explicit form of the monomials $W_i$ and their coefficients.]{\label{tab:Wi}
\rule{0cm}{2em}Explicit form of the monomials $W_i$ and their coefficients.}
\endlastfoot
\multicolumn{1}{|r|}{$1$} & \multicolumn{1}{|l|}{$i\left\langle \left[
S\cdot u,v\cdot u\right] h_{\mu }^{\;\mu }\right\rangle $} & 
\multicolumn{1}{|l|}{$\frac{1}{16}\left( g_{A}+g_{A}^{3}\right) +\frac{3}{4}%
g_{A}mc_{4}$} \\ 
\multicolumn{1}{|r|}{$2$} & \multicolumn{1}{|l|}{$\left\langle
h_\mu^{\;\mu} h_\nu^{\;\nu} \right\rangle$}
& \multicolumn{1}{|l|}{$-\frac{1}{2} m c_2$} \\ 
\multicolumn{1}{|r|}{$3$} & \multicolumn{1}{|l|}{$v^{\mu }v^{\nu
}\left\langle h_{\mu \nu }h_{\rho }^{\;\rho }\right\rangle $} & 
\multicolumn{1}{|l|}{$-\frac{1}{16}g_{A}^{2}+\frac{1}{2}m^{2}\left(
d_{14}-d_{15}\right) $} \\ 
\multicolumn{1}{|r|}{$4$} & \multicolumn{1}{|l|}{$\left\langle
u_\mu [D^\mu,h_\nu^{\;\nu}]\right\rangle$}
& \multicolumn{1}{|l|}{$-2 m c_2$} \\ 
\multicolumn{1}{|r|}{$5$} & \multicolumn{1}{|l|}{$\left\langle v\cdot u\left[
v\cdot D,h_{\mu }^{\;\mu }\right] \right\rangle $} & \multicolumn{1}{|l|}{$-%
\frac{1}{16}g_{A}^{2}-m^{2}\left(d_{14}-d_{15}\right)$} \\ 
\multicolumn{1}{|r|}{$6$} & \multicolumn{1}{|l|}{$v^{\mu }\langle u^{\nu }%
\left[ v\cdot D,h_{\mu \nu }\right] $} & \multicolumn{1}{|l|}{$-\frac{1}{8}%
g_{A}^{2}+m^{2}\left( d_{14}-d_{15}\right) $} \\ 
\multicolumn{1}{|r|}{$7$} & \multicolumn{1}{|l|}{$v^{\mu }v^{\nu
}\left\langle v\cdot u\left[ v\cdot D,h_{\mu \nu }\right] \right\rangle $} & 
\multicolumn{1}{|l|}{$\frac{1}{4}g_{A}^{2}$} \\ 
\multicolumn{1}{|r|}{$8$} & \multicolumn{1}{|l|}{$[S^{\mu },S^{\nu }]v^{\rho
}\left[ u_{\nu },\left[ v\cdot D,h_{\mu \rho }\right] \right] $} & 
\multicolumn{1}{|l|}{$\frac{1}{4}\left( 1+g_{A}^{2}\right) +mc_{4}$} \\
\hline
\end{longtable}

\renewcommand{\arraystretch}{1.2}
\setlength{\LTcapwidth}{\textwidth}

\begin{longtable}[c]{|c|c|c|}  
\hline
$i$ & $X_{i}^{\lambda }$ & $8m^{3}e_{i}^{\prime \prime }$ \\ \hline
\hline
\hline
\endhead
\hline
\caption[]{\rule{0cm}{2em}}
\endfoot
\hline
\caption[Explicit form of the monomials $X_i^\lambda$ and their coefficients.]{\label{tab:Xi}
\rule{0cm}{2em}Explicit form of the monomials $X_i^\lambda$ and their coefficients.}
\endlastfoot
\multicolumn{1}{|r|}{$1$} & \multicolumn{1}{|l|}{$i\langle S\cdot uv\cdot
u\rangle u^{\lambda }$} & \multicolumn{1}{|l|}{$\frac{1}{4}\left(
g_{A}+g_{A}^{3}\right) +2g_{A}mc_{4}+4m^{2}d_{13}$} \\ 
\multicolumn{1}{|r|}{$2$} & \multicolumn{1}{|l|}{$iS\cdot u\langle v\cdot
uu^{\lambda }\rangle $} & \multicolumn{1}{|l|}{$-\frac{1}{4}\left(
g_{A}+g_{A}^{3}\right) -2g_{A}mc_{4}+8m^{2}d_{12}$} \\ 
\multicolumn{1}{|r|}{$3$} & \multicolumn{1}{|l|}{$iS^{\lambda }\langle
u\cdot u\rangle v\cdot u$} & \multicolumn{1}{|l|}{$\frac{1}{8}\left(
2g_{A}+g_{A}^{3}\right) -g_{A}m\left( c_{3}-c_{4}\right) -4m^{2}d_{10}$} \\ 
\multicolumn{1}{|r|}{$4$} & \multicolumn{1}{|l|}{$iS^{\lambda }\langle
(v\cdot u)^{2}\rangle v\cdot u$} & \multicolumn{1}{|l|}{$-\frac{1}{8}%
g_{A}^{3}-g_{A}mc_{2}-4m^{2}(d_{12}+d_{13})$} \\ 
\multicolumn{1}{|r|}{$5$} & \multicolumn{1}{|l|}{$iS^{\lambda }\langle
v\cdot uu_{\mu }\rangle u^{\mu }$} & \multicolumn{1}{|l|}{$-\frac{1}{4}%
g_{A}-g_{A}mc_{4}-4m^{2}d_{11}$} \\ 
\multicolumn{1}{|r|}{$6$} & \multicolumn{1}{|l|}{$i\langle S\cdot
uu^{\lambda }\rangle v\cdot u$} & \multicolumn{1}{|l|}{$4m^{2}d_{13}$} \\ 
\multicolumn{1}{|r|}{$7$} & \multicolumn{1}{|l|}{$iS\cdot u \langle
(v\cdot u)^{2}\rangle v^{\lambda }$} & \multicolumn{1}{|l|}{$\frac{3}{8}%
\left( 2g_{A}+g_{A}^{3}\right) +3g_{A}mc_{4}$} \\ 
\multicolumn{1}{|r|}{$8$} & \multicolumn{1}{|l|}{$i\langle S\cdot u
v\cdot u\rangle v\cdot uv^{\lambda }$} & \multicolumn{1}{|l|}{$-\frac{1}{4}%
\left( 3g_{A}+2g_{A}^{3}\right) -3g_{A}mc_{4}$} \\
\hline
\end{longtable}

\renewcommand{\arraystretch}{1.2}

\begin{longtable}[c]{|c|c|c|}  
\hline
$i$ & $Y_{i}^{\mu \nu }$ & $8m^{3}e_{i}^{\prime \prime \prime }$ \\ \hline
\hline
\hline
\endhead
\hline
\caption[]{\rule{0cm}{2em}}
\endfoot
\hline
\caption[Explicit form of the monomials $Y_i^{\mu\nu}$ and their coefficients.]{\label{tab:Yi}
\rule{0cm}{2em}Explicit form of the monomials $Y_i^{\mu\nu}$ and their coefficients.}
\endlastfoot
\multicolumn{1}{|r|}{$1$} & \multicolumn{1}{|l|}{$\left\langle u^\mu u^\nu \right\rangle$} & 
\multicolumn{1}{|l|}{$-2 m c_2$} \\ 
\multicolumn{1}{|r|}{$2$} & \multicolumn{1}{|l|}{$g^{\mu \nu }\left\langle
u\cdot u\right\rangle $} &
\multicolumn{1}{|l|}{$\frac{1}{16}g_{A}^{2}-\frac{1}{2}mc_{3}
\hspace{3cm} $
} \\ 
\multicolumn{1}{|r|}{$3$} & \multicolumn{1}{|l|}{$v^{\mu }\left\langle
u\cdot u\right\rangle v^{\nu }$} & \multicolumn{1}{|l|}{$-\frac{1}{16}%
g_{A}^{2}+\frac{1}{2} mc_{3}$} \\ 
\multicolumn{1}{|r|}{$4$} & \multicolumn{1}{|l|}{$g^{\mu \nu }\left\langle
\left( v\cdot u\right) ^{2}\right\rangle $} & \multicolumn{1}{|l|}{$-\frac{1%
}{16}g_{A}^{2}-\frac{1}{2} mc_{2}$} \\ 
\multicolumn{1}{|r|}{$5$} & \multicolumn{1}{|l|}{$v^{\mu }\left\langle
\left( v\cdot u\right) ^{2}\right\rangle v^{\nu }$} & \multicolumn{1}{|l|}{$%
\frac{3}{8}g_{A}^{2}+\frac{1}{2} mc_{2}$} \\ 
\multicolumn{1}{|r|}{$6$} & \multicolumn{1}{|l|}{$g^{\mu \nu }\left[ S^{\rho
},S^{\tau }\right] \left[ u_{\rho },u_{\tau }\right] $} & 
\multicolumn{1}{|l|}{$-\frac{1}{8}g_{A}^{2}-\frac{1}{2} mc_{4}$} \\ 
\multicolumn{1}{|r|}{$7$} & \multicolumn{1}{|l|}{$v^{\mu }\left[ S^{\rho
},S^{\tau }\right] \left[ u_{\rho },u_{\tau }\right] v^{\nu }$} & 
\multicolumn{1}{|l|}{$-\frac{1}{8}\left( 1-g_{A}^{2}\right) +\frac{1}{2} mc_{4}$} \\ 
\multicolumn{1}{|r|}{$8$} & \multicolumn{1}{|l|}{$\left\langle u^{\mu
}v\cdot u\right\rangle v^{\nu }$} & \multicolumn{1}{|l|}{$-\frac{1}{4}%
g_{A}^{2}$} \\
\hline
\end{longtable}

\begin{longtable}[c]{|c|c|c|}  
\hline
$i$ & $Z_{i}^{\lambda \mu \nu }$ & $8m^{3}e_{i}^{iv}$ \\ \hline
\hline
\hline
\endhead
\hline
\caption[]{\rule{0cm}{2em}}
\endfoot
\hline
\caption[Explicit form of the monomials $Z_i^{\lambda\mu\nu}$  
         and their coefficients.]{\label{tab:Zi}
\rule{0cm}{2em}Explicit form of the monomials $Z_i^{\lambda\mu\nu}$  
         and their coefficients.}
\endlastfoot
\multicolumn{1}{|r|}{$1$} & \multicolumn{1}{|l|}{$iu^{\lambda }S^{\mu
}v^{\nu }$} & \multicolumn{1}{|l|}{$-2g_{A}$} \\ 
\multicolumn{1}{|r|}{$2$} & \multicolumn{1}{|l|}{$iS\cdot ug^{\lambda \mu
}v^{\nu }$} & \multicolumn{1}{|l|}{$g_{A}$} \\ 
\multicolumn{1}{|r|}{$3$} & \multicolumn{1}{|l|}{$iv\cdot uS^{\lambda
}v^{\mu }v^{\nu }$} & \multicolumn{1}{|l|}{$3g_{A}$} \\ 
\multicolumn{1}{|r|}{$4$} & \multicolumn{1}{|l|}{$iS\cdot uv^{\lambda
}v^{\mu }v^{\nu }$} & \multicolumn{1}{|l|}{$-g_{A}$} \\ \hline
\end{longtable}

\medskip\noindent
We briefly comment on the fourth order heavy baryon Lagrangian
recently presented in ref.\cite{M99}. First, it is entirely based
on the HB projection and contains no information concerning the
relativistic formulation. Furthermore, strong isospin breaking terms
as well as external sources (apart from the quark mass matrix)
are entirely omitted. The tables in that work contain considerably
more terms than found here, which could be reduced by some of the
identities spelled out in this paper.

\section{Summary}

In this work, we have constructed the minimal effective
chiral pion--nucleon two--flavor  Lagrangian at fourth order in the chiral
expansion. To arrive at this minimal form, we have studied in detail
all strictures arising from the equations of motion, trace relations
and other algebraic identities. The so constructed Lagrangian contains
118 in principle measurable  terms. These are accompanied by the so--called
low--energy constants, which can be obtained by a fit to data, from
some models or maybe from lattice gauge theory. Note that four of these
operators are special since they have no pion matrix elements.
Based on the fact
that a consistent power counting can be set up in the relativistic
as well as in the heavy baryon formulation of the effective field
theory, we have worked out the effective Lagrangian in both schemes.
In the heavy baryon case, it is mandatory to calculate the various
$1/m$ corrections at a given chiral order. Here, these are explicitly 
spelled out up--to--and--including  fourth order.

\subsection*{Acknowledgment}

We would like to thank Miro Helbich and Guido M\"uller for 
collaboration in the early stages of the work, 
Thomas Becher for a useful comment, and Peter Stelmachovic for 
pointing out a typographical error.

\appendix
\def\theequation{\Alph{section}.\arabic{equation}}
\setcounter{equation}{0}
\section{EOM eliminations}
\label{app:EOM}

The number of independent terms in the effective $\pi$N chiral Lagrangian is
considerably reduced by the use of the so-called EOM (equation of motion)
eliminations. One can view them simply as replacements of $\barr{D}\Psi$ by $-im\Psi$
(a consequence of the fact that $\left(\barr{D}+im\right) \Psi$ is of higher order), or
as a consequence of the lowest order EOM (defining the classical solution,
around which the loop expansion is performed), or as a result of  specific
field transformations (leaving the S--matrix elements untouched).
However, the simple replacements $\barr{D}\Psi\rightarrow-im\Psi$ do
not exhaust  the EOM
eliminations. The point is that also terms (or combinations of terms) not
containing $\barr{D}\Psi$ explicitly can be brought into a form that
includes $\barr{D}\Psi$. Such terms can then also be eliminated. To
systematically obtain all such terms, one has to
inspect all the possible terms with $\Gamma_{\ldots}\barr{D}\Psi$ and rewrite them as
combinations of terms without $\barr{D}\Psi$.
To achieve this, we expand any product $\Gamma_{\ldots}\gamma_{\lambda}$ in the
form (the last $\gamma$--matrix originates from $\barr{D}$)
\begin{equation}
\Gamma_{\ldots}\gamma_{\lambda}=\sum_{i}\Gamma_{i,\ldots\lambda}^{\prime}
+\sum_{j}\Gamma_{j,\ldots\lambda}^{\prime\prime}~, \qquad c_{\Gamma
_{j}^{\prime\prime}}=c_{\Gamma_{i}^{\prime}}+1=c_{\Gamma}~,
\end{equation}
where $\Gamma_{i}^{\prime}$ and $\Gamma_{j}^{\prime\prime}$ contain only
elements from the standard basis of the Clifford algebra, multiplied by
appropriate combinations of $g$- and $\varepsilon$-tensors. When contracted
with $D_{\lambda}$ and inserted into (\ref{monom2}) instead of $\Gamma
_{\ldots}$, this yields (after the replacement $\barr{D}\Psi\rightarrow-im\Psi$)
\begin{eqnarray}
\sum_{i}\bar{\Psi}A^{\mu\nu\ldots\alpha\beta\ldots}\Gamma_{i,\mu\nu
\ldots}^{\prime\qquad\lambda}D_{\lambda\alpha\beta\ldots}^{n}\Psi
+\mathrm{h.c.} & \doteq & -im\bar{\Psi}A^{\mu\nu\ldots\alpha\beta\ldots
}\Gamma_{\mu\nu\ldots}D_{\alpha\beta\ldots}^{n-1}\Psi+\mathrm{h.c.}~,\nonumber\\
\sum_{j}\bar{\Psi}A^{\mu\nu\ldots\alpha\beta\ldots}\Gamma_{j,\mu\nu
\ldots}^{\prime\prime\qquad\lambda}D_{\lambda\alpha\beta\ldots}^{n}%
\Psi+\mathrm{h.c.} & \doteq & 0~,
\label{eom gen}%
\end{eqnarray}
where the symbol $\doteq$ stands for ''equal up-to higher orders''.

\medskip\noindent
For $\Gamma=1$, one obtains  the so--called ``master relation''
\begin{equation}
\bar{\Psi}A^{\alpha\beta\ldots}\gamma^{\lambda}D_{\lambda\alpha
\beta\ldots}^{n}\Psi+\mathrm{h.c.}\doteq-im\bar{\Psi}A^{\alpha\beta
\ldots}D_{\alpha\beta\ldots}^{n-1}\Psi+\mathrm{h.c.}~, \label{eom}%
\end{equation}
but letting $\Gamma$ run through the rest of the standard basis of the
Clifford algebra, a couple of additional relations is generated
\begin{equation}
0\doteq-im\bar{\Psi}A^{\alpha\beta\ldots}\gamma_{5}D_{\alpha\beta\ldots
}^{n-1}\Psi+\mathrm{h.c.}~, \label{eom1}%
\end{equation}%
\begin{equation}
\bar{\Psi}A^{\alpha\beta\ldots}\gamma^{5}\gamma^{\lambda}D_{\lambda
\alpha\beta\ldots}^{n}\Psi+\mathrm{h.c.}\doteq0~, \label{eom2}%
\end{equation}%
\begin{equation}
\bar{\Psi}A^{\mu\alpha\beta\ldots}D_{\mu\alpha\beta\ldots}^{n}%
\Psi+\mathrm{h.c.}\doteq-im\bar{\Psi}A^{\mu\alpha\beta\ldots}\gamma_{\mu
}D_{\alpha\beta\ldots}^{n-1}\Psi+\mathrm{h.c.}~, \label{eom3}%
\end{equation}%
\begin{equation}
\bar{\Psi}A^{\mu\alpha\beta\ldots}\sigma_{\mu}^{\;\lambda}D_{\lambda
\alpha\beta\ldots}^{n}\Psi+\mathrm{h.c.}\doteq0~, \label{eom4}%
\end{equation}%
\begin{equation}
\bar{\Psi}A^{\mu\alpha\beta\ldots}\gamma_{5}\sigma_{\mu}^{\;\lambda
}D_{\lambda\alpha\beta\ldots}^{n}\Psi+\mathrm{h.c.}\doteq m\overline{\Psi
}A^{\mu\alpha\beta\ldots}\gamma_{5}\gamma_{\mu}D_{\alpha\beta\ldots}^{n-1}%
\Psi+\mathrm{h.c.}~, \label{eom5}%
\end{equation}%
\begin{equation}
\bar{\Psi}A^{\mu\alpha\beta\ldots}\gamma_{5}D_{\mu\alpha\beta\ldots}%
^{n}\Psi+\mathrm{h.c.}\doteq0~, \label{eom6}%
\end{equation}%
\begin{equation}
\bar{\Psi}A^{\mu\nu\alpha\beta\ldots}\varepsilon_{\mu\nu}^{\quad
\lambda\rho}\gamma_{5}\gamma_{\rho}D_{\lambda\alpha\beta\ldots}^{n}%
\Psi+\mathrm{h.c.}\doteq im\bar{\Psi}A^{\mu\nu\alpha\beta\ldots}%
\sigma_{\mu\nu}D_{\alpha\beta\ldots}^{n-1}\Psi+\mathrm{h.c.}~, \label{eom7}%
\end{equation}%
\begin{equation}
\bar{\Psi}A^{\mu\nu\alpha\beta\ldots}\left(  \gamma_{\mu}D_{\nu
\alpha\beta\ldots}^{n}-\gamma_{\nu}D_{\mu\alpha\beta\ldots}^{n}\right)
\Psi+\mathrm{h.c.}\doteq0~. \label{eom8}%
\end{equation}
For the sake of completeness, we have written down both relations (\ref{eom gen}) for
every $\Gamma$ $\in$ $(\gamma_{5}$,
$\gamma_{\mu}$,$\gamma_{5}\gamma_{\mu}$, $\sigma_{\mu
\nu})$, in spite of the fact that they are not all independent (e.g.\ eqs.(\ref{eom1})
and (\ref{eom6}) are equivalent, and (\ref{eom8}) follows from (\ref{eom3})).

\medskip\noindent
For a general $\Gamma$ one obtains, as a rule, some linear combinations of the
above relations. But there are exceptions, namely for $\varepsilon$--tensors
contracted with elements of the standard basis. The first such case is
$\Gamma=\gamma_{5}\sigma_{\mu\nu}=(1/2i)\varepsilon_{\mu\nu\rho\tau
}\sigma^{\rho\tau}$, which yields two new relations,
\begin{equation}
\bar{\Psi}A^{\mu\nu\alpha\beta\ldots}\left(  \gamma_{5}\gamma_{\mu
}D_{\nu\alpha\beta\ldots}^{n}-\gamma_{5}\gamma_{\nu}D_{\mu\alpha\beta\ldots
}^{n}\right)  \Psi+\mathrm{h.c.}
\doteq-m\bar{\Psi}A^{\mu\nu\alpha\beta\ldots}\gamma_{5}\sigma_{\mu\nu
}D_{\alpha\beta\ldots}^{n-1}\Psi+\mathrm{h.c.}~, \label{eom9}
\end{equation}
\begin{equation}
\bar{\Psi}A^{\mu\nu\alpha\beta\ldots}\varepsilon_{\mu\nu}^{\quad
\lambda\rho}\gamma_{\rho}D_{\lambda\alpha\beta\ldots}^{n}\Psi+\mathrm{h.c.}%
\doteq0~. \label{eom10}%
\end{equation}
For $\Gamma=\varepsilon_{\mu\nu\rho\tau}\gamma^{\tau}$, one obtains
\begin{equation}
\bar{\Psi}A^{\mu\nu\rho\alpha\ldots}\varepsilon_{\;\mu\nu\rho}^{\lambda
}D_{\lambda\alpha\ldots}^{n}\Psi+\mathrm{h.c.}\doteq-im\bar{\Psi}%
A^{\mu\nu\rho\alpha\ldots}\varepsilon_{\;\mu\nu\rho}^{\lambda}\gamma_{\lambda
}D_{\alpha\ldots}^{n-1}\Psi+\mathrm{h.c.}~, \label{eom11}%
\end{equation}%
\begin{equation}
\bar{\Psi}A^{\mu\nu\rho\alpha\ldots}\gamma_{5}\left(  g_{\mu}^{\lambda
}\sigma_{\nu\rho}+g_{\rho}^{\lambda}\sigma_{\mu\nu}+g_{\nu}^{\lambda}%
\sigma_{\rho\mu}\right)  D_{\lambda\alpha\ldots}^{n}\Psi+\mathrm{h.c.}%
\doteq0~, \label{eom12}%
\end{equation}
and $\Gamma=\varepsilon_{\mu\nu\rho\tau}\gamma_{5}\gamma^{\tau}$ gives
\begin{align}
& \bar{\Psi}A^{\mu\nu\rho\alpha\ldots}\left(  g_{\mu}^{\lambda}%
\sigma_{\nu\rho}+g_{\rho}^{\lambda}\sigma_{\mu\nu}+g_{\nu}^{\lambda}%
\sigma_{\rho\mu}\right)  D_{\lambda\alpha\ldots}^{n}\Psi+\mathrm{h.c.}
\nonumber \\
& \doteq im\bar{\Psi}A^{\mu\nu\rho\alpha\ldots}\varepsilon_{\mu\nu\rho
}^{\quad\,\, \tau}\gamma_{5}\gamma_{\tau}D_{\alpha\ldots}^{n-1}\Psi+\mathrm{h.c.}~,
\label{eom13}
\end{align}%
\begin{equation}
\bar{\Psi}A^{\mu\nu\rho\alpha\ldots}\gamma_{5}\varepsilon_{\;\mu\nu\rho
}^{\lambda}D_{\lambda\alpha\ldots}^{n}\Psi+\mathrm{h.c.}\doteq0~. \label{eom14}%
\end{equation}
The relations (\ref{eom1}--\ref{eom14}) can be used to eliminate some terms
from the list of all the possible invariants, but it is preferable to use
them already at the stage of generating such a list.\footnote{In \cite{Krause}
and \cite{BK98} implications of the nucleon EOM have been expressed in the form of
six identities. The first five of them are equivalent to each other, and they
are consequences of our relations (\ref{eom1}), (\ref{eom3}--\ref{eom11}) and
(\ref{eom14}). The sixth relation of \cite{BK98} follows from our relations
(\ref{eom}) and (\ref{eom2}), and is not given explicitly (although is
accounted for) by \cite{Krause}. The sixth relation of \cite{Krause} is neither
given, nor used by \cite{BK98}. The relation follows from our relations
(\ref{eom12}--\ref{eom13}), however, not for a general $A^{\mu\nu\rho\ldots}$,
but rather only for $A^{\mu\nu\rho\ldots}$ antisymmetric in $\mu\nu$ (or
$\mu\rho$ or $\nu\rho$).} This was exactly what we did, e.g.\ when counting
$\gamma_{5}$ as a quantity of first chiral order in Table~\ref{signs1}, or
when considering the special structure of $\Theta_{\mu\nu\ldots}$ in
eq.(\ref{monom}). In this way, the relations (\ref{eom}--\ref{eom2}) and
(\ref{eom4}--\ref{eom6}) were taken into account. The relations (\ref{eom3}),
(\ref{eom8}) and (\ref{eom11}) are also easily accounted for, namely by ignoring all
structures $\Theta_{\mu\nu\ldots}$ with an explicit $\gamma_{\mu}$
(but not with  an explicit
$\gamma_{5}\gamma_{\mu}$). The relations (\ref{eom7}) and 
(\ref{eom9}--\ref{eom10}) are simultaneously embodied in the simple requirement that
$\Theta_{\mu\nu\ldots}$ can contain an $\varepsilon$--tensor with at most one
index contracted within $\Theta_{\mu\nu\ldots}$. The relation (\ref{eom12}) is
just a simple consequence of eq.(\ref{eom9}), and the relation (\ref{eom13}) allows to
ignore $\Theta_{\mu\nu\rho\ldots}$ containing 
$\varepsilon_{\mu\nu\rho\tau}\gamma_{5}\gamma^{\tau}$.

\medskip\noindent
So far we have investigated the implications of the nucleon EOM on the structure
of $\Theta_{\mu\nu\ldots}$. There are, however, also consequences for the
structure of $A^{\mu\nu\ldots}$, namely for $A^{\mu\nu\ldots}=\left[  D^{\mu
},B^{\nu\ldots}\right]  $. The point is that $D^{\mu}$ can be reshuffled to
act on nucleon fields (by partial integration and exclusion of total
derivatives) and, in some cases, eliminated afterwards. Such an elimination is
obviously possible if $D^{\mu}$ is contracted with $\gamma_{\mu}D_{\nu\ldots
}^{n}$ and consequently also if it is contracted with $\gamma_{\nu}%
D_{\mu\ldots}^{n}$ (cf. eq.(\ref{eom8})) or $D_{\mu\nu\ldots}^{n}$ 
(cf. eq.(\ref{eom3})). For $D^{\mu}$ contracted with $\gamma_{5}\gamma_{\mu}D_{\nu
\ldots}^{n}$ and $\gamma_{5}\gamma_{\nu}D_{\mu\ldots}^{n} $ one obtains in a
similar way
\begin{equation}
\bar{\Psi}\left[  D^{\mu},B^{\nu\ldots}\right]  \gamma_{5}\gamma_{\mu
}D_{\nu\ldots}^{n}\Psi+\mathrm{h.c.}\doteq2im\bar{\Psi}B^{\nu\ldots
}\gamma_{5}D_{\nu\ldots}^{n}\Psi+\mathrm{h.c.}~,\label{DB1}%
\end{equation}%
\begin{equation}
\bar{\Psi}\left[  D^{\mu},B^{\nu\ldots}\right]  \gamma_{5}\gamma_{\nu
}D_{\mu\ldots}^{n}\Psi+\mathrm{h.c.}\doteq0~, \label{DB2}%
\end{equation}
and for $D^{\mu}$ contracted with $\sigma_{\nu\rho}D_{\mu\ldots}^{n}$
\begin{eqnarray}%
\bar{\Psi}\left[  D^{\mu},B^{\nu\rho\ldots}\right]  \sigma_{\nu\rho
}D_{\mu\ldots}^{n}\Psi+\mathrm{h.c.} & \doteq &  i\bar{\Psi}\left[
D^{\mu},B^{\nu\rho\ldots}\right]  g_{\mu\nu}D_{\rho\ldots}^{n}\Psi
+\mathrm{h.c.}\nonumber\\
& - & i\bar{\Psi}\left[  D^{\mu},B^{\nu\rho\ldots}\right]  g_{\mu\rho
}D_{\nu\ldots}^{n}\Psi+\mathrm{h.c.}\nonumber\\
& + & im\bar{\Psi}\left[  D^{\mu},B^{\nu\rho\ldots}\right]
\varepsilon_{\mu\nu\rho}^{\quad\;\tau}\gamma_{5}\gamma_{\tau}D_{\ldots}%
^{n-1}\Psi+\mathrm{h.c.}~~.%
\label{DB3}%
\end{eqnarray}
Note that the omitted contraction of $D^{\mu}$ with $\sigma_{\mu\nu}D_{\rho\ldots}%
^{n}$ yields nothing but a trivial identity.
The relations (\ref{DB1}--\ref{DB3}) are used to ignore  terms of the
type written on the left--hand--side. However,
in case of eq.(\ref{DB1}) it is more advantageous to keep the
structure on the left--hand--side and eliminate the right--hand--side,
which does not contain the special structure $\left[  D^{\mu},B^{\nu\ldots
}\right]  $, but rather a general structure $B^{\nu\ldots}$.

\medskip\noindent
To summarize: we utilize the nucleon EOM in the form of restrictions on the
structure of $\Gamma_{\mu\nu\ldots}$ and $A^{\mu\nu\ldots}$:
\begin{itemize}
\item $\Gamma_{\mu\nu\ldots}$ is a product of $g$-tensors, $\varepsilon
$-tensors and one matrix from the following set $\{1,\gamma_{5}\gamma_{\mu
},\sigma_{\mu\nu}\}$,
\item  within $\Gamma_{\mu\nu\ldots}$, only $\varepsilon$-tensors may have an
upper index (every $\varepsilon$ at most one), which has to be contracted with
$D_{\alpha\beta\ldots}^{n}$,
\item  for $A^{\mu\nu\ldots}=\left[  D^{\mu},B^{\nu\ldots}\right]  $,
derivative $D^{\mu}$ cannot be contracted with $D_{\mu\alpha\ldots}^{n}$
\end{itemize}
\noindent This is not the only way of implementing the nucleon EOM, but it
seems to be the most economic one. It results in the following set of
$\Theta_{\mu\nu\ldots}$, sufficient for generating the complete chiral
Lagrangian up to fourth order:
\begin{eqnarray}%
&&1;\nonumber\\
&&\gamma_{5}\gamma_{\mu},\;D_{\mu};\nonumber\\
&&g_{\mu\nu},\;\sigma_{\mu\nu},\;\gamma_{5}\gamma_{\mu}D_{\nu},\;D_{\mu\nu};\nonumber\\
&&g_{\mu\nu}\gamma_{5}\gamma_{\rho},\;g_{\mu\nu}D_{\rho},\;\sigma_{\mu\nu
}D_{\rho},\;\varepsilon_{\;\mu\nu\rho}^{\lambda}D_{\lambda},\;\gamma_{5}%
\gamma_{\mu}D_{\nu\rho},\;D_{\mu\nu\rho};\nonumber\\
&&g_{\mu\nu}g_{\rho\tau},\;\varepsilon_{\mu\nu\rho\tau},\;g_{\mu\nu}\sigma
_{\rho\tau},\;g_{\mu\nu}\gamma_{5}\gamma_{\rho}D_{\tau},\;g_{\mu\nu}%
D_{\rho\tau},\;\sigma_{\mu\nu}D_{\rho\tau},\;\nonumber\\
&&\varepsilon_{\;\mu\nu\rho}^{\lambda}D_{\lambda\tau},\;\gamma_{5}\gamma_{\mu
}D_{\nu\rho\tau},\;D_{\mu\nu\rho\tau}\, .%
\end{eqnarray}

\def\theequation{\Alph{section}.\arabic{equation}}
\setcounter{equation}{0}
\section{Operators for the heavy baryon Lagrangian}
\label{app:LHB}

In order to construct the $1/m$-corrections up-to-and-including fourth
order, one needs the expressions for the operators 
${\cal A}^{(1,2,3,4)}$, ${\cal B}^{(1,2,3)}$, and ${\cal C}^{(0,1,2)}$, the
latter allowing one to reconstruct the chiral expansion of the inverse
of ${\cal C}$. These are given by:
\begin{eqnarray}
{\cal C}^{(0)}  & = & 2 m~, \\
{\cal A}^{(1)} & = & i\, (v\cdot D) + g_A (S\cdot u)~,\\
{\cal B}^{(1)} & = & -\gamma_5\,\left(2 i \, (S\cdot D) 
+\frac{g_A}{2} (v\cdot u) \right)~,\\
{\cal C}^{(1)} & = & i\,(v\cdot D) + g_A (S\cdot u)~,
\end{eqnarray}
\begin{eqnarray}
{\cal A}^{(2)} & = & c_1 \lanl \chi_+ \ranl +\frac{c_2}{2} \lanl (v\cdot u)^2\ranl
+\frac{c_3}{2} \lanl u^2\ranl + c_5 \widetilde{\chi}_+ \nonumber\\
&-&2 i\, [S^\mu,S^\nu] \left( \frac{i c_4}{4} [u_\mu,u_\nu] + \frac{c_6}{8 m} F_{\mu\nu}^+
+\frac{c_7}{8 m} \lanl F_{\mu\nu}^+\ranl \right)~, \\
{\cal B}^{(2)} & = & 2 i \gamma_5 (v^\mu S^\nu - v^\nu S^\mu)
\left( \frac{i c_4}{4} [u_\mu,u_\nu] + \frac{c_6}{8 m} F_{\mu\nu}^+
+\frac{c_7}{8 m} \lanl F_{\mu\nu}^+\ranl \right)~, \\
{\cal C}^{(2)} & = & 
- {\cal A}^{(2)}~, 
\\
{\cal A}^{(3)} & = &  \Big( i \,\,\frac{c_2}{2 m} \lanl (v\cdot u) u_\mu \ranl \, D^\mu 
+ {\rm h.c.} \Big)+  E + 2 S^\mu G_\mu - 2 i\,[S^\mu,S^\nu] H_{\mu\nu}~, \\
{\cal B}^{(3)} & = & \gamma_5 \left( -v^\mu G_\mu 
+ 2 i\,(v^\mu S^\nu - v^\nu S^\mu) H_{\mu\nu} \right)~,
\\
{\cal A}^{(4)}& = & -\frac{c_2}{8 m^2} \lanl u_\mu u_\nu \ranl 
                                \{D^\mu,D^\nu\} + {\rm h.c.}\no\\
&&- \frac{d_1}{2m} \Big( [u_\mu,[D_\nu,u^\mu]] D^\nu 
                       + \mbox{h.c.} 
                 \Big)
- \frac{d_2}{2m} \Big( [u_\mu,[D^\mu,u_\nu]] D^\nu 
                       + \mbox{h.c.} 
                 \Big)
\nonumber \\ & &
-\,\frac{d_3}{2 m} \Big( [v\cdot u,[v\cdot D,u_\mu]] D^\mu  
                         +[v\cdot u,[D_\mu,v\cdot u]] D^\mu  
                         +[u_\mu,[v\cdot D,v\cdot u]] D^\mu  + \mbox{h.c.} 
                     \Big)
\nonumber \\ & &
-\,\frac{d_4}{2m} \Big( \epsilon^{\mu\nu\alpha\beta} 
                        \langle u_\mu u_\nu u_\alpha \rangle D_\beta
                        + \mbox{h.c.} 
                  \Big)
+ \frac{d_5}{2m} \Big(i\, [\chi_-,u_\mu] D^\mu
                          + \mbox{h.c.} 
                    \Big)
\nonumber \\ & &
+\,\frac{d_6}{2m} \Big(i\, [D^\mu,\widetilde{F}^+_{\mu\nu}] D^\nu
                           + \mbox{h.c.} 
                     \Big)
+ \frac{d_7}{2m} \Big( i\,[D^\mu,\langle F^+_{\mu\nu}\rangle] D^\nu 
                          + \mbox{h.c.} 
                    \Big)
\nonumber \\ & &
+\,\frac{d_8}{2m} \Big( i\,\epsilon^{\mu\nu\alpha\beta} 
                           \langle \widetilde{F}^+_{\mu\nu} u_\alpha \rangle D_\beta
                           + \mbox{h.c.} 
                     \Big)
+ \frac{d_9}{2m} \Big( i\,\epsilon^{\mu\nu\alpha\beta} 
                          \langle F^+_{\mu\nu} \rangle u_\alpha D_\beta
                          + \mbox{h.c.} 
                    \Big)
\nonumber \\ & &
+ \frac{d_{12}}{m} \Big(i\,\langle v\cdot u u_\mu \rangle
                            S\cdot u D^\mu
                            + \mbox{h.c.} 
                      \Big)
\nonumber \\ & &
+\,\frac{d_{13}}{2 m} \Big(i\, S^\mu \langle u_\mu v\cdot u \rangle u_\nu D^\nu
                            + i\,S^\mu \langle u_\mu u_\nu \rangle v\cdot u D^\nu 
                            + \mbox{h.c.}
                       \Big)
\nonumber \\ & &
+ \frac{d_{14}}{2m} \Big( [S^\mu,S^\nu]
                             \langle [D_\lambda,u_\mu] u_\nu \rangle D^\lambda
                             + \mbox{h.c.} 
                       \Big)
\nonumber \\ & &
+\,\frac{d_{15}}{2m} \Big( [S^\mu,S^\nu]
                              \langle u_\mu [D_\nu,u_\lambda] \rangle D^\lambda
                              + \mbox{h.c.} 
                        \Big)
\nonumber \\ & &
-\,\frac{d_{20}}{2m} \Big(S^\mu v^\nu[\widetilde{F}^+_{\mu\nu},u_\lambda]D^\lambda 
                         +S^\mu [\widetilde{F}^+_{\mu\nu},v\cdot u]D^\nu+\mbox{h.c.} 
                          \Big)~,
\end{eqnarray}
with
\begin{eqnarray}
E & = &   i d_1  [u_\mu,[v\cdot D,u^\mu]]  
        + i d_2  [u_\mu,[D^\mu,v\cdot u]] 
        + i d_3  [v\cdot u,[v\cdot D,v\cdot u]] \nonumber \\
        &+& i d_4  \epsilon^{\mu\nu\alpha\beta} v_\beta\lanl u_\mu u_\nu u_\alpha \ranl
        + d_5  [\chi_-,v\cdot u ] 
        + d_6  v^\nu  [D^\mu,\widetilde{F}^+_{\mu\nu}] \nonumber \\ 
        &+& d_7  v^\nu  [D^\mu,\langle F^+_{\mu\nu}\rangle]
        + d_8  \epsilon^{\mu\nu\alpha\beta} v^\beta
                           \langle \widetilde{F}^+_{\mu\nu} u_\alpha \rangle 
        + d_9  \epsilon^{\mu\nu\alpha\beta} v^\beta 
                          \langle F^+_{\mu\nu} \rangle u_\alpha~,  \\
G_\mu & = & \frac{d_{10}}{2} \langle u^2 \rangle u_\mu
+ \frac{d_{11}}{2} \langle u_\mu u_\nu \rangle u^\nu 
+ \frac{d_{12}}{2}  \langle (v\cdot u)^2 \rangle
                            u_\mu 
+ \frac{d_{13}}{2}  \langle u_\mu (v\cdot u) \rangle
                             (v\cdot u) \nonumber\\
&&+ \frac{d_{16}}{2} \langle \chi_+ \rangle u_\mu
+ \frac{d_{17}}{2} \langle \chi_+ u_\mu \rangle 
+\,\frac{i\,d_{18}}{2} [D_\mu,\chi_-]
+ \frac{i\,d_{19}}{2} [D_\mu,\langle \chi_- \rangle]
+\,\frac{i\,d_{20}}{2} v^\nu  [\widetilde{F}^+_{\mu\nu},v\cdot u] \nonumber \\
&&+ \frac{i\,d_{21}}{2} [\widetilde{F}^+_{\mu\nu},u^\nu]
+\,\frac{d_{22}}{2} [D^\nu,F^-_{\mu\nu}]
+ \frac{d_{23}}{2} \epsilon_\mu^{\,\,\,\nu\alpha\beta} 
                   \langle u_\nu F^-_{\alpha\beta} \rangle~, \\
H_{\mu\nu} & = & 
\frac{d_{14}}{2} \langle [v\cdot D,u_\mu] u_\nu \rangle 
+\,\frac{d_{15}}{2}  \langle u_\mu [D_\nu,v\cdot u] \rangle~,
\end{eqnarray}
and
\begin{eqnarray}
{\cal C}^{-1} & = & \frac{1}{2 m}
-\frac{i v\cdot D + g_A S\cdot u}{(2 m)^2}
+\frac{(i v\cdot D + g_A S\cdot u)^2}{(2 m)^3} 
- \frac{{\cal C}^{(2)}}{(2 m)^2}~.
\end{eqnarray}


\end{document}